\documentclass[12pt,twoside,leqno]{article}
\newtheorem{Def}{Definition}[section]
\newtheorem{Pro}{Proposition}[section]
\newtheorem{Teo}{Theorem}[section]
\newtheorem{Lem}{Lemma}[section]

\newtheorem{Rem}{Remark}[section]
\newenvironment{proof}[1][Proof]{\textbf{#1.} }{\ \hfill{\fbox{}}}
\begin{document}
\title{\hspace{6cm} \underline{\small{Harmonic Analysis}}\\Spectral analysis on Damek-Ricci space}
\author{By\\ Ahmed Abouelaz\\Depart.Mathematics,Faculty of Sciences Ain-chock,Casablanca-Morocco\\E.mail:
aboulaz$@$facsc-achok.ac.ma\\ and\\ Omar El Fourchi\\ Amal Idafi
5,N° 258 Hay el Fath, Rabat-Morocco\\E.mail:
elfourchi$\_$omar$@$yahoo.fr}
\date{}
\maketitle \pagestyle{myheadings}
\markboth{\underline{\centerline{\textit{\small{Spectral analysis
on Damek-Ricci
space}}}}}{\underline{\centerline{\textit{\small{Spectral analysis
on Damek-Ricci space}}}}} {\bf Abstract.}--- We define and study
the spectral projection operator for compactly supported
distributions on Damek-Ricci space $NA$. The Paley-Wiener-Schwartz
theorem and the range of ${\mathcal{S}}^{p}(NA)^{\#}$($0<p\leq 2$)
via spectral projection operator are established. The
$L^{2}$-estimation for this operator is also given. In order to do
the Paley-Wiener theorem for the non necessary radial function,
the spectral projection operator can be uniquely characterized by
analyticity and growth condition in $\lambda$ of Paley-Wiener
theorem type on the unit disk of the complex plane as an example
of Damek-Ricci space. \footnotetext[1]{1991 Mathematics subject
classification. 42B38, 44A15, 92C55.} \footnotetext[2]{Key Words
and phrases. Paley-Wiener Theorem, Spectral projection operator,
Damek-Ricci spaces.}

\section{Introduction}
Given a group $N$ of Heisenberg type, let $S=NA$ be the
one-solvable extension of $N$ obtained by letting $A= I\!\!R^{+}$
acts on $N$ by homogeneous dilatation. We equip $S$  with a
natural left-invariant Riemannian structure. The group $S$ is in
generally nonsymmetric harmonic spaces (the geodesic symmetry
around the identity is not isometry (see [13] and [15])). The
geodesic distance of $x\in NA $ from the identity e is
\begin{eqnarray*}
\rho(x)= d(x,e)= \log(\frac{1+r(x)}{1-r(x)}),\hspace{1cm}  0\le
r(x)\le 1
\end{eqnarray*}
with,
\begin{eqnarray*}
r(V,Z,a)^2 =1-\frac{4a}{(1+a+ \frac{|V|^2}{4})^2 +|Z|^2}.
\end{eqnarray*}
On such a group $S$, we consider the Laplacian $\pounds $ on $NA$
whose radial part is given by the rule
\begin{eqnarray*}
\pounds_r
=\frac{\partial^2}{\partial\rho^2}+(\frac{m}{2}\coth(\frac{\rho}{2})+k\coth
(\rho))\frac{\partial}{\partial\rho},
\end{eqnarray*}
with $\rho$ is the geodesic distance of $x\in NA$ from the the
identity. The Fourier-Helgason transform of the function $f$ in
${\mathcal{D}}$(NA) (see [7]) is the function $\widehat{f}$ on
$l\!\!\!C\times N$ defined by
\begin{eqnarray}
\widehat{f}(\lambda,n)=\int_{NA}f(x){\mathcal{P}}_{\lambda}(x,n)\,dx
,
\end{eqnarray}
where the kernel ${\mathcal{P}}_{\lambda}$ : $NA \times N
\longrightarrow l\!\!\!C$ is an appropriate complex power of the
poisson kernel on $NA$, namely \begin{eqnarray*}
{\mathcal{P}}_{\lambda}(x,n)=[{\mathcal{P}}(x,n)]^{\frac{1}{2}-i\frac{\lambda}{Q}}.
\end{eqnarray*}
For a distribution, the Fourier-Helgason transform of  $T\in
{\mathcal{E}}'$(NA) (see [1] and [2]) is  defined by
\begin{eqnarray}
\widehat{T}(\lambda,n)=\langle
T(x),{\mathcal{P}}_{\lambda}(x,n)\rangle
\hspace{0.5cm}(\lambda,n)\in l\!\!\!C \times N,
\end{eqnarray}
the above formula have a sens, because the function $x\to
{\mathcal{P}}_{\lambda}(x,n)$ is an eigenfunction of
Laplace-Beltrami operator with eigenvalue
$-(\lambda^{2}+\frac{Q^2}{4})$ (see [6] and [7]).\\ This
transform, for $f$ in ${\mathcal{D}}$(NA), can be also written as
follows (see [2])
\begin{eqnarray}
\widehat{f}(\lambda,n)= {\mathcal{F}}_{1}\circ R_{n}f(\lambda),
\end{eqnarray}
where ${\mathcal{F}}_{1}$ is the Fourier transform of one variable
and $R_{n}f(\lambda)$ is the horocyclic Radon transform (see [2])
defined as follows,
\begin{eqnarray}
R_{n}f(\lambda)=e^{-\lambda \frac{Q}{2}}\int_{N}f(n\sigma
(n_{1}\exp(\lambda H)))\,dn_{1},
\end{eqnarray}
with $\sigma$ is the geodesic inversion (see [8]) defined by the
following formula
\begin{eqnarray*}
\sigma(V,Z,t)=
\frac{1}{(t+\frac{|V|^2}{4})^2+|Z|^2}[((-t+\frac{|V|^2}{4})+J_Z)V,-Z,t],
\end{eqnarray*}
for all $(V,Z,t)\in NA$, and write the Fourier-Helgason inversion
formula (see [7])
\begin{eqnarray}
f(x)=\int_{-\infty}^{+\infty}(\frac{c_{m,k}}{4\pi}|c(\lambda)|^{-2}\int_{N}{\mathcal{P}}_{-\lambda}(x,n)\widehat{f}(\lambda,n)\,dn
)\,d{\lambda},
\end{eqnarray}
where
$c_{m,k}=2^{k-1}\Gamma(\frac{2m+k+1}{2})\frac{1}{\pi^(\frac{2m+k+1}{2})}$
and  $c(\lambda)$ is the generalized Harish-chadra function.\\ The
expression in Parentheses in the formula (1.5) is an
eigenfunctions of Laplace-Beltrami operator with eigenvalue
$-(\lambda^{2}+\frac{Q^2}{4})$.\\ For $f\in {\mathcal{D}}(NA)$, We
define the spectral projector as follows
\begin{eqnarray}
I\!\!P_{\lambda}f(x)=\frac{c_{m,k}}{4\pi}|c(\lambda)|^{-2}\int_{N}{\mathcal{P}}_{-\lambda}(x,n)\widehat{f}(\lambda,n)\,dn.
\end{eqnarray}
so that the spectral representation is
\begin{eqnarray}
f(x)=\int_{-\infty}^{+\infty}I\!\!P_{\lambda}f(x)\,d\lambda.
\end{eqnarray}
For a distribution, the spectral projection of  $T\in
{\mathcal{E}}'$(NA) is defined as follows
\begin{eqnarray}
I\!\!PT(x)=\frac{c_{m,k}}{4\pi}|c(\lambda)|^{-2}(T*\phi_{\lambda}(x)),
\end{eqnarray}
 The above formula have a sens, because $\phi_{\lambda}$ is the spherical function, which is ${\mathcal{C}}^{\infty}(NA)$.\\
\\
{\bf Remark.} On the unit open disk $D=\{z\in l\!\!\!C; |z|=1\}$
of the complex plane, the Laplace-Beltrami operator $\Delta_{D}$
on $D$ (see [20]) can be written in terms of the Euclidean
Laplacian $\Delta_{I\!\!R^2}$ as
\begin{eqnarray*}
\Delta_{D}&=&(1-|z|^{2})^{2}\Delta_{I\!\!R^2}\\
&=&4(1-|z|^{2})^{2}\frac{\partial^{2}}{\partial z\partial
\bar{z}}.
\end{eqnarray*}
For $z\in D$, let $r, \theta \in I\!\!R$, with $r\geq 0$ be such
that $z=\tanh (r)  e^{i\theta}$. Since $d(0,z)$=r, then
$(r,\theta)$ are called geodesic polar coordinates of $z$. In such
coordinates of $z$, the Laplace-Beltrami is
\begin{eqnarray*}
\Delta_{D}=\frac{\partial^{2}}{\partial r^{2}}+
2coth(2r)\frac{\partial}{\partial
r}+4\sinh^{-2}(2r)\frac{\partial^{2}}{\partial \theta^{2}}
\end{eqnarray*}
For the Laplacian $\Delta_{D}$ on $D$ (see [20]), we have
\begin{eqnarray*}
\Delta_{D}
(e^{(i\lambda+1)<z,w>})=-(\lambda^{2}+1)e^{(i\lambda+1)<z,w>},\lambda
\in l\!\!\!C.
\end{eqnarray*}
With
\begin{eqnarray*}
e^{<z,w>}=(\frac{1-|z|^{2}}{|1-
z.\overline{w}|^{2}})^{\frac{1}{2}}.
\end{eqnarray*}
For $\lambda \in l\!\!\!C$, let ${\mathcal{P}}_{\lambda}$ denote
the complex power of the Poisson kernel ( cf. [20] p: 3) given by
\begin{eqnarray*}
{\mathcal{P}}_{\lambda}(z,w)&=&e^{(i\lambda+1)<z,w>}\\
&=&(\frac{1-|z|^{2}}{|1-
z.\overline{w}|^{2}})^{\frac{i\lambda+1}{2}},
\end{eqnarray*}
One can define the Fourier-Helgason transform by
\begin{eqnarray*}
(1.8)'\,\,\widehat{f}(\lambda,w)=\int_{D}{\mathcal{P}}_{-\lambda}(z,w)f(z)\,d\mu(z),
\end{eqnarray*}
for all $\lambda \in l\!\!\!C$, $w\in S^1$ for which this integral
exists, where $d\mu(z)=(1-|z|^2)^{-2}dz$, and the Fourier-Helgason
inversion formula is(cf. [20] p: 33)
\begin{eqnarray*}
f(z)=\frac{1}{4\pi}\int_{I\!\!R}(\int_{S^{1}}\widehat{f}(\lambda,w){\mathcal{P}}_{\lambda}(z,w)\,d\sigma(w))\lambda
\tanh(\frac{\pi \lambda}{2})\,d\lambda.
\end{eqnarray*}
The expression in parentheses is an eigenfunction of the Laplacian
on $D$ with eigenvalue $-(\lambda^{2}+1)$, we define the spectral
projection operator on the $D$ (see [20]) as follows
\begin{eqnarray*}
(1.8)''\,\,\,I\!\!P_{\lambda}f(z)&=&\frac{1}{4\pi}\lambda
\tanh(\frac{\pi
\lambda}{2})\int_{S^{1}}\widehat{f}(\lambda,w){\mathcal{P}}_{\lambda}(z,w)\,d\sigma(w)\\
&=&\frac{1}{4\pi}\lambda \tanh(\frac{\pi
\lambda}{2}){\mathcal{Q}}_{\lambda}f(z).
\end{eqnarray*}
{\bf Remark}. We note that $I\!\!P_{\lambda}f(z)$ is not defined
at $\lambda_{l}=\frac{+}{}i(2k+1)$ for $k \in l\!\!Z^{*}$, and has
a simple zero at a points $\lambda_{h}=\frac{+}{}i 2 h $ for $h
\in l\!\!Z^{*}$ and a double zeros at $\lambda=0$. Also the
function $\lambda \to {\mathcal{Q}}_{\lambda}f(z)$ is even. \\
{\bf Remark}. For $k\in l\!\!Z^{+}$, we have
$${\mathcal{Q}}_{-i(2k+1)}f(z)=(1-|z|^2)^{-k}\int_{S^{1}}\widehat{f}(-i(2k+1),w)(1-z.\overline{w})^{2k}\,d\sigma(w).$$
Using the Libnitz formula to the function
$(1-z.\overline{w})^{2k}$, we note that the the integrale in the
second part of the above formula is a polynmial of $z$ and
$\overline{z}$ of degree $2k$.\\
\\
The aim of this work is to characterize the range of
${\mathcal{E}}'(NA)^{\#}$ (respectively of
${\mathcal{S}}^{p}(NA)^{\#}$ for $0<p\leq 2$) by the spectral
projection operator (see Theorem 3.2) (respectively Theorem 4.1),
we find the analogous of this theorem in the case of the
noncompact symmetric space of rank one (see [26] and [27]), and
mainly we give an estimation of this operator in $L^{2}(NA)$,
also, we discuss in the sense of R. Strichartz (cf. [26]), the
spectral Paley-Wiener theorem on the unit open disk of the complex
plane .\\
\\
Now we give a full description of the organization of this paper.
In section 2, we recall the main definition and the know results
of spherical analysis on NA groups. In section 3, we introduce the
Poisson kernel and spectral projection, we state the main results,
we have obtained the characterization of the
${\mathcal{E}}'(NA)$-range of the spectral projection which is a
generalization of theorem 3.6 in [26](see also [27]). In section
4, we give  a range of ${\mathcal{S}}^{p}(NA)^{\#}$ (for $0<p\leq
2$) via spectral projection. In section 5, we discut the
$L^{2}$-estimate for this projection (see theorem 5.1 and 5.2). In
the next, we discuss in the sense of R. Strichartz (cf. [26]), the
spectral Paley-Wiener theorem on the unit open disk $D$ of the
complex plane as an example of the hyperbolic spaces (even case).
the results and ideas will be  illustrated by developing a range
theorem for spectral projection operator on $D$. And mainly to
characterize the
${\mathcal{C}}_{com}^{\infty}(\overline{B_R(z_0)})$-range (where
$B_R(z_0)$ is the unit ball of $l\!\!\!C$ centered at $z_0$) of
spectral projection operator $I\!\!P_{\lambda}$ associated to the
Laplacian $\Delta_D$ on $D$.
\section{Notations and Preliminaries}
Let \underline{$\eta$} be a two-step real nilpotent Lie algebra of
finite dimensional (i.e.,  [\underline{$\eta$},\underline{$\eta$}]
$\neq 0$ and
[\underline{$\eta$},[\underline{$\eta$},\underline{$\eta$}]]=0)
equipped with an inner product $<,>$, \underline{$\eta$} has a
center \underline{z}. We have then $[V,V']\in \underline{z}$ and
[V,Z]=0  $\forall V,V'\in \underline{\eta}$ and  $\forall Z\in
\underline{z}$. We write \underline{$\eta$} as an orthogonal sum
of two spaces \underline {p} and \underline{z}
(\underline{$\eta$}=\underline{p}$\oplus$\underline{z}, we have
$[\underline{p},\underline{p}]\subset \underline{z}$,
$[\underline{p},\underline{z}]=0$ and
$[\underline{z},\underline{z}]=0$. According to Kaplan [21],
\underline{$\eta$} is said to be an H-type Lie algebra if for
every unitary $Z\in \underline{z}$ the map $J_Z$ of \underline{p}
into \underline{p}, defined by equality
$<J_{Z}V,V'>_{\underline{\eta}}=<[V,V'],Z>_{\underline{\eta}}$,
satisfy the equality $J_{Z}^{2}V= -|Z|^{2}V$, for all $V\in
\underline {p}$. A fondamental example is the Heisenberg algebra
(see [24]), given by the matrix. $$\left (\begin{array}{ccccc}
0&v_1&...&v_k&z\\
 &   &   &   &w_1\\
 &   &(0)&   &. \\
 &   &   &   &w_k\\
 &   &   &   &0
\end{array}
\right)=(v,w,z),\hspace{0,5cm}v,w \in I\!\!R^k ,z\in I\!\!R$$ such
that $J_{(0,0,z)}(v,w,0)=z(-w,v,0)$.\\ Note that for every unit
$Z\in \underline{z}$, $J_{Z}$ is a complex structure on
$\underline{p}$, so that $\underline{p}$ has even dimension m=2m',
we denote by k the dimension of \underline{z}. Let N be the
connected and simply connected group of Lie algebra
\underline{$\eta$}. Since  \underline{$\eta$} is nilpotent, the
exponential map is surjective, we may therefore parametrize N by
\underline{p}$\oplus$\underline{z} and write (V,Z) for
$\exp{(V+Z)}$  where $V\in ${\underline{p}} and $Z\in$
{\underline{z}}. By the Baker-Campbell-Hausdorff formula, the
product law in $N$ is given by the formula
\begin{eqnarray*}
(V,Z).(V',Z')=(V+V',Z+Z'+\frac{1}{2}[V,V']),
\end{eqnarray*}
for all $V,V'\in \underline{p}$ and for all $Z,Z'\in
\underline{z}$. Let $dV$ and $dZ$ the lebesgue measures on
$\underline{p}$ and $\underline{z}$
 respectively, the measure $dVdZ$ is the Haar measure on $N$  whose we denote
by $dn$. Let $A$ be a multiplicatif group  isomorphe to
$I\!\!R_{+}^{*}$  and  $NA$ the semi-direct product of $N$ and $A$
relatively to the action $(V,Z)\in{\underline{\eta}} \longmapsto
(a^{\frac{1}{2}}V,aZ)$. So the Lie group $S=NA$ (connected and
simply connected) is called a Damek-Ricci space. We denote by
$(V,Z,a)$ the element $na =\exp(V+Z)a$, the inner law on the group
$NA$ is given by the formula
\begin{eqnarray*}
(V,Z,a).(V',Z',a')=(V+a^{\frac{1}{2}}V',Z+aZ'+\frac{1}{2}a^{\frac{1}{2}}[V,V'],aa').
\end{eqnarray*}
Denote by $Q=\frac{1}{2}m+k$, with $Q=2\varrho$  the homogeneous
dimension of $N$, the left Haar measure on $NA$ is given by
$dx=a^{-Q-1}dVdZda=a^{-Q-1}dnda$. Note that the right Haar measure
on $NA$ is $a^{-1}dVdZda$, then the group $NA$ is nonunimodulaire,
so that the modular function  $\delta $ is given by
$\delta(V,Z,a)=a^{-Q}$. As Riemannian manifold , $NA$ is (see
[15]) a harmonic space, the noncompact symmetric space of rank one
is contained in these class of $NA$ groups , $ NA\approx G/K
=NAK/K$ ( here $NA$ is the Iwasawa group). Also the group $S$
provide an examples of nonsymmetric hamonic spaces (see [15]). The
eigenfunction may be expressed as Jacobi of parameters $\alpha$
and $\beta$ via the following formula (see [22] p. 152)
\begin{eqnarray}
\Phi_{s}(x)&=&\Phi_{s}(\rho)\\ \nonumber
               &=&\varphi_{2\lambda}^{(\alpha,\beta)}(\frac{\rho}{2})\\      &=&_{2}F_1(\frac{1}{2}(Q-2s),\frac{1}{2}(Q+2s);\frac{m+k+1}{2};-\sinh^2(\frac{\rho}{2})).\nonumber
\end{eqnarray}
where recall that $_{2}F_1$ is the Gauss hypergeometric function
with $\alpha =\frac{m+k-1}{2}$, $\beta =\frac{k-1}{2}$ and
$\lambda =-iRe(s) +Im(s)$, then $Im(\lambda) = -Re(s)).$ \\
\section{Poisson kernel and Spectral Projection on the Damek-Ricci space}
For $n_1$ fixed in $N$, we define (see [7] p. 409), the Poisson
kernel on $NA$ for $n_1$ by the formula
\begin{eqnarray*}
{\mathcal{P}}(.,n_1): NA &\longrightarrow& I\!\!R \\ na &\mapsto
&{\mathcal{P}}(na,n_1)=P_{a} (n_{1}^{-1}n),
\end{eqnarray*}
where, for $a>0$, $P_{a}(n)$ is a function on $N$ defined by
\begin{eqnarray}
P_{a}(n)=P_{a}(V,Z)=a^Q((a+\frac{|V|^2}{4})^2+|Z|^2)^{-Q}.
\end{eqnarray}
We have the following properties
\begin{itemize}
\item $\pounds {\mathcal{P}}(.,n_1)= 0,\hspace{1cm} \forall n_{1}\in N$
\item $P_{a}(n)=a^{-Q}P_1(a^{-1}na),\hspace{1cm} \forall a\in A, \forall n\in N.$
\end{itemize}
With these properties, one may defined the kernel
${\mathcal{P}}_{\lambda}$, $(\lambda\in l\!\!\!C)$ on $NA\times N$
as follows $$
\begin{array}{cccccc}
{\mathcal{P}}_{\lambda}:&NA\times N&\longrightarrow &l\!\!\!C &
&\\
&(na,\overline{n})&\longrightarrow&{\mathcal{P}}_{\lambda}(na,\overline{n})&=&
{\mathcal{P}}(na,\overline{n})^{\frac{1}{2}-\frac{i\lambda}{Q}}\\
& & &
&=&P_{a}(\overline{n}^{-1}n)^{\frac{1}{2}-\frac{i\lambda}{Q}}
\end{array}
$$ we define the spectral projection operator on the Damek-Ricci
space and we study these properties
\begin{Def}
Let $T$ be an element of ${\mathcal{E}}'(NA)$, we define the
spectral projection operator on NA as follows
\begin{eqnarray}
I\!\!P_{\lambda}T(x)=\frac{c_{m,k}}{4\pi}|c(\lambda)|^{-2}(T*\Phi_{\lambda})(x).
\end{eqnarray}
\end{Def}
\begin{Pro}
Let $T$ be an element of ${\mathcal{E}}'(NA)$, then for all
$\lambda \in l\!\!\!C$ we have
\begin{eqnarray}
I\!\!P_{\lambda}T(x)=\frac{c_{m,k}}{4\pi}|c(\lambda)|^{-2}\int_{N}{\mathcal{P}}_{-\lambda}(x,n)\widehat{T}(\lambda,n)\,dn.
\end{eqnarray}
\end{Pro}
\begin{proof} Let $T\in {\mathcal{E}}'(NA)$, from the formula 3.2, we obtain
\begin{eqnarray}
I\!\!P_{\lambda}T(x)&=&\frac{c_{m,k}}{4\pi}|c(\lambda)|^{-2}\langle
T(y),\Phi_{\lambda}(y^{-1}x)) \rangle\\
&=&\frac{c_{m,k}}{4\pi}|c(\lambda)|^{-2}\langle
T(y),\Phi_{\lambda}(d(x,y)) \rangle.\nonumber
\end{eqnarray}
The spherical function $\Phi_{\lambda}$ satisfies to the following
formula (see [24] p. 42 and [7] p. 413)
\begin{eqnarray}
\Phi_{\lambda}(x^{-1}y)=\int_{N}{\mathcal{P}}_{-\lambda}(x,n){\mathcal{P}}_{\lambda}(y,n)\,dn.
\end{eqnarray}
Using the Fubini-theorem, the formula (3.4) becomes
\begin{eqnarray}
I\!\!P_{\lambda}T(x)&=&\frac{c_{m,k}}{4\pi}|c(\lambda)|^{-2}\langle
T(y),\int_{N}{\mathcal{P}}_{-\lambda}(x,n){\mathcal{P}}_{\lambda}(y,n)\,dn
\rangle\\ &=&
\frac{c_{m,k}}{4\pi}|c(\lambda)|^{-2}\int_{N}{\mathcal{P}}_{-\lambda}(x,n)\langle
T,{\mathcal{P}}_{\lambda}(y,n) \rangle \,dn \nonumber\\ &=&
\frac{c_{m,k}}{4\pi}|c(\lambda)|^{-2}\int_{N}{\mathcal{P}}_{-\lambda}(x,n)\widehat{T}(\lambda,n)\,dn
\nonumber
\end{eqnarray}
\end{proof}\\
\\
Let $M$ : ${\mathcal{D}}(NA)\to {\mathcal{D}}(NA)^{\#}$ be the
averaging projector on NA (see [7], [14] and [24]) defined as
follows
\begin{eqnarray*}
(Mf)(x)&=&\frac{1}{|{\mathcal{S}}|}\int_{{\mathcal{S}}_{\rho}}f(y)\,d\sigma_{\rho}(y),\\
\end{eqnarray*}
where $d\sigma_{\rho}$ is the surface measure induced by the
left-invariant Riemannian metric on the geodesic sphere
${\mathcal{S}}_{\rho}=\{y\in NA : d(y,e)=\rho\}$, normalised by
$\int_{{\mathcal{S}}_{\rho}}\,d\sigma_{\rho}(y)=1$ and
$\rho(x)=d(x,e)$. Denote by $f_{x}$ the function
$M(\tau_{x^{-1}}f)$  where $x,y$ $\in$ NA and
$\tau_{x}g(y)=g(x^{-1}y)$ is the translated function.
\begin{Pro} Let x$\in NA$ and f be in  ${\mathcal{D}}(NA)$ , then
\begin{eqnarray} I\!\!P_{\lambda}f(x)=\frac{c_{m,k}}{4\pi}|c(\lambda)|^{-2}\widetilde{f_{x}}(\lambda)
\end{eqnarray}
where $\widetilde{f}$ design the spherical Fourier transform of f
$\in {\mathcal{D}}(NA)^{\#}$,
\end{Pro}
\begin{proof} Let $x$ be an element of $NA$ and $f\in {\mathcal{D}}(NA)$ , since $f_{x}$ is a radial function on NA, the spherical Fourier transform of $f_{x}$  is given by
\begin{eqnarray*}
\widetilde{f_{x}}(\lambda)&=&\int_{NA}f_{x}(y)\Phi_{\lambda}(y)\,dy\\
&=&\int_{NA}f(x^{-1}y)\Phi_{\lambda}(y)\,dy,
\end{eqnarray*}
putting $x^{-1}y=z$ to obtain
\begin{eqnarray*}
\widetilde{f_{x}}(\lambda)=f*\Phi_{\lambda}(x),
\end{eqnarray*}
and this prove the proposition.
\end{proof}\\
\\
Remark that for $x=e$ and  $f\in {\mathcal{D}}(NA)^{\#}$, the
equality (3.7) becomes
\begin{eqnarray}
I\!\!P_{\lambda}f(e)=\frac{c_{m,k}}{4\pi}|c(\lambda)|^{-2}\widetilde{f}(\lambda).
\end{eqnarray}
In order to do the Paley-Wiener theorem for the spectral
projection operator, we will need the following lemma.
\begin{Lem}(Koornwinder (see [22] p. 150))
For each $\alpha,\beta \in l\!\!\!C$ and for each non-negative
integer n there exists a positive constant $C$ such that for all
$t\geq 0$ and all $\lambda \in l\!\!\!C$
\begin{eqnarray*}
|(\Gamma(\alpha
+1))^{-1}\frac{d^{n}}{dt^{n}}\varphi_{\lambda}^{(\alpha,\beta)}(t)|\leq
C(1+|\lambda|)^{n+k}(1+t)e^{(|Im\lambda|-Re\mu)t},
\end{eqnarray*}
where  $\mu=\alpha +\beta +1$, $k=0$ if Re$\alpha >-\frac{1}{2}$
and $k=\frac{1}{2}-Re\alpha$ if $Re\alpha \leq -\frac{1}{2}$,
where the function $t\to \varphi_{\lambda}^{(\alpha,\beta)}(t)$ is
the Jacobi function.
\end{Lem}
Let $B_{a}(z)$ be the ball of center $z\in NA$ and of radius a for
the distance $d$. We denote by
${\mathcal{C}}_{c}^{\infty}(B_{a}(z))$ the set of function $f\in
{\mathcal{D}}(NA)$ which suppf is inclued in $B_{a}(z)$
\begin{Lem}
If $f\in {\mathcal{C}}_{c}^{\infty}(B_{a}(z))\bigcap
{\mathcal{D}}(NA)^{\#}$, then $I\!\!P_{\lambda}f(x)$ satisfies to
following conditions:\\ 1) For all $\lambda \in 1\!\!\!C$, the
function $x\to I\!\!P_{\lambda}f(x)$ is a radial function\\
2)$(\lambda,x)\to I\!\!P_{\lambda}f(x)$ is a
${\mathcal{C}}^{\infty}$ function on $1\!\!\!C \times NA$\\ 3)
for all $\lambda \in 1\!\!\!C$, we have
${\mathcal{L}}_rI\!\!P_{\lambda}f(x)=-(\lambda^{2}+\varrho^{2})I\!\!P_{\lambda}f(x)$
(where ${\mathcal{L}}_r$ is the radial part of the
Laplace-Beltrami operator)\\ 4)  for each fixed x, the function
$I\!\!P_{\lambda}f(x)$ is an entire function divisible by
$|c(\lambda)|^{-2}$ and the quotient is an analytic function\\ 5)
for every $N_0$ there exists $C_{N_0}$ such that
\begin{eqnarray*}
|I\!\!P_{\lambda}f(x)|\leq
C_{N_0}|c(\lambda)|^{-2}(1+|\lambda|^{2})^{-N_0}e^{|Im
\lambda|(d(x,z)+a)}
\end{eqnarray*}
\end{Lem}
{\bf Remark.} the above theorem hold for all dimension of $NA$\\
\\
\begin{proof}
The spherical function $\Phi_{\lambda}(x)$ is given by the formula
(see [6]).
\begin{eqnarray*}
\Phi_{\lambda}(x)&=&\int_{N}{\mathcal{P}}_{-\lambda}(x,n){\mathcal{P}}_{\lambda}(e,n)\,dn
\\ &=&\int_{N}e^{(\varrho+i\lambda)(t\circ
\sigma)(n^{-1}x)+(\varrho-i\lambda)(t\circ \sigma)(n^{-1})}\,dn
\end{eqnarray*}
and the equality (6.6) shows that $(\lambda,x)\to
I\!\!P_{\lambda}f(x)$ is a ${\mathcal{C}}^{\infty}$ function on
$1\!\!\!C \times NA$. The formula (6.6)implies, also, that
${\mathcal{L}}I\!\!P_{\lambda}f(x)=-(\lambda^{2}+\varrho^{2})I\!\!P_{\lambda}f(x)$
because $\Phi_{\lambda}$ is an eigenfunction for ${\mathcal{L}}_r$
and the operator ${\mathcal{L}}$ has for eigenvalue
$-(\lambda^{2}+\varrho^{2})$. It follows from (6.6) that
$I\!\!P_{\lambda}f(x)$ is even ($x$ fixed) since
$\Phi_{\lambda}=\Phi_{-\lambda}$, the equality 6.6 shows that
$I\!\!P_{\lambda}f(x)$ is divisible by $|c(\lambda)|^{-2}$.
Showing, now, the condition 5). Assume that supp $f$ is included
in the ball $B_{a}(z)$ ($z$ fixed) and let ${\mathcal{L}}_{0}$ be
the operator defined by
${\mathcal{L}}_{0}=-{\mathcal{L}}+\varrho^{2}$ with $\varrho=2Q$,
where ${\mathcal{L}}_r$ is the radial part of the Laplace Beltrami
operator (see the introduction). Let $r$ be the integers, then
\begin{eqnarray}
I\!\!P_{\lambda}({\mathcal{L}}_{0}^{r}f)(x)=(-1)^{r}\lambda^{2r}I\!\!P_{\lambda}f(x).
\end{eqnarray}
But , from (6.6) we have
\begin{eqnarray*}
I\!\!P_{\lambda}({\mathcal{L}}_{0}^{r}f)(x)&=&\frac{c_{m,k}}{4\pi}|c(\lambda)|^{-2}({\mathcal{L}}_{0}^{r}f*\Phi_{\lambda})(x)\\
&=&\frac{c_{m,k}}{4\pi}|c(\lambda)|^{-2}\int_{NA}\Phi_{\lambda}(xy^{-1})({\mathcal{L}}_{0}^{r})f(y)\,dy.
\end{eqnarray*}
By (6.18), the above equality becomes
\begin{eqnarray*}
(-1)^{r}\lambda^{2r}I\!\!P_{\lambda}f(x)=\frac{c_{m,k}}{4\pi}|c(\lambda)|^{-2}\int_{NA}\Phi_{\lambda}(xy^{-1})({\mathcal{L}}_{0}^{r})f(y)\,dy.
\end{eqnarray*}
This equality implies
\begin{eqnarray}
|\lambda|^{2r}|I\!\!P_{\lambda}f(x)|&\leq&
\frac{c_{m,k}}{4\pi}(|c(\lambda)|^{-2}\sup_{y\in
NA}|{\mathcal{L}}_{0}^{r}f(y)|).\\ &&\times(\int_{y\in
B_{a}(z)}|\Phi_{\lambda}(xy^{-1})|\,dy)\nonumber
\end{eqnarray}
It follows from the Koornwinder lemma (see lemma 6.1) that
\begin{eqnarray}
|\lambda|^{2r}|I\!\!P_{\lambda}f(x)|&\leq&
\frac{c_{m,k}}{4\pi}(|c(\lambda)|^{-2})(\sup_{y\in
NA}|{\mathcal{L}}_{0}^{r}f(y)|).\\ &&\times(|B_{a}(z)|)\sup_{y\in
B_{a}(z)}e^{|Im \lambda|\rho(xy^{-1})}\,dy.\nonumber
\end{eqnarray}
Where $|B_{a}(z)|$ design the measure of the ball $B_{a}(z)$.
Since
$\rho(xy^{-1})=d(x,y) \leq d(x,z)+d(z,y) \leq d(x,z)+a$ ,
the inequality (6.19) can be transformed as follows
\begin{eqnarray*}
|I\!\!P_{\lambda}f(x)|\leq
c'_{r}|c(\lambda)|^{-2}(1+|\lambda|^{2})^{-r}e^{|Im
\lambda|(d(x,z)+a)}.
\end{eqnarray*}
Where $c'_{r}$ is an other absolute constant. The third condition
of the above lemma is lawful because the condition (6.13) implies
that
$\frac{4\pi}{c_{m,k}}(I\!\!P_{\lambda}f(e)|c(\lambda)|^{2})=\widetilde{f}(\lambda)$
for a radial function $f$. According to the theorem 3.14 in [16],
we known that the function $\lambda \to \widetilde{f}(\lambda)$ is
analytic.
\end{proof}
\begin{Teo} (Abouelaz see [1]) .
If $f\in {\mathcal{C}}_{c}^{\infty}(B_{a}(z))\bigcap
{\mathcal{D}}(NA)^{\#}$, then $I\!\!P_{\lambda}f(x)$ satisfies to
following conditions:\\ 1) For all $\lambda \in 1\!\!\!C$, the
function $x\to I\!\!P_{\lambda}f(x)$ is a radial function\\
2)$(\lambda,x)\to I\!\!P_{\lambda}f(x)$ is a
${\mathcal{C}}^{\infty}$ function on $1\!\!\!C \times NA$\\ 3)
for all $\lambda \in 1\!\!\!C$, we have
${\mathcal{L}}I\!\!P_{\lambda}f(x)=-(\lambda^{2}+\varrho^{2})I\!\!P_{\lambda}f(x)$
(where ${\mathcal{L}}$ is the radial part of the Laplace-Beltrami
operator)\\ 4)  for each fixed x, the function
$I\!\!P_{\lambda}f(x)$ is an entire function divisible by
$|c(\lambda)|^{-2}$ and the quotient is an analytic function\\ 5)
for every $N_0$ there exists $C_{N_0}$ such that
\begin{eqnarray*}
|I\!\!P_{\lambda}f(x)|\leq
C_{N_0}|c(\lambda)|^{-2}(1+|\lambda|^{2})^{-N_0}e^{|Im
\lambda|(d(x,z)+a)}
\end{eqnarray*}
Conversely, if $x\to F(\lambda,x)$ (for all $\lambda \in
l\!\!\!\!C$)is a radial function and $F(\lambda,x)$ satisfies to
1),2),3),4) and 5) then there exist $f\in
{\mathcal{C}}_{c}^{\infty}(B_{a}(z))\bigcap
{\mathcal{D}}(NA)^{\#}$ such that
$I\!\!P_{\lambda}f(x)=F(\lambda,x)$ for all $(\lambda,x) \in
l\!\!\!\!C \times NA)$ Where ${\mathcal{L}}$ is the radial part of
the Laplace-Beltrami operator (see (2.2))
\end{Teo}

\begin{proof}
The necessary condition is proved in the above lemma.\\
conversely,let $F(\lambda,x)$ be a function which satisfy to
condition (1),..,(4), the condition (5) of the above theorem shows
that
\begin{eqnarray*}
(a_1)\,\,|F(\lambda,x)|\leq
C_{N_0}|c(\lambda)|^{-2}(1+|\lambda|^{2})^{-N_0}e^{|Im
\lambda|(d(x,z)+a)}.
\end{eqnarray*}
Without loss of generality we take z=e (as in the proof of theorem
of R. Strichartz in [26]).  The function $x\to F(\lambda,x)$ (for
$\lambda fixed$) is radial verifying the equality
\begin{eqnarray*}
(a_2)\hspace{0.5cm}{\mathcal{L}}F(\lambda,x)=-(\lambda^{2}+\rho^{2})F(\lambda,x).
\end{eqnarray*}
The function
\begin{eqnarray*}
\Psi(\lambda,x)=\frac{F(\lambda,x)}{F(\lambda,e)}.
\end{eqnarray*}
verify the equality $(a_2)$, then
\begin{eqnarray*}
F(\lambda,x)=F(\lambda,e)\Phi_{\lambda}(x).
\end{eqnarray*}
(where $\Phi_{\lambda}(x)$ is the spherical function). Replace
$F(\lambda,x)$ by its expression in the equality $(a_1)$, we
obtain
\begin{eqnarray*}
(a_3)\hspace{0.3cm}|F(\lambda,e)||\Phi_{\lambda}(x)|\leq
C_{N_0}|c(\lambda)|^{-2}(1+|\lambda|^{2})^{-N_0}e^{|Im
\lambda|(r+a)}.
\end{eqnarray*}
(where $r=\rho(x)$. The inequality $(a_3)$ implies that
\begin{eqnarray*}
(a_4)\hspace{0.3cm}|F(\lambda,e)|e^{-|Im
\lambda|r}|\Phi_{\lambda}(x)|\leq
C_{N_0}|c(\lambda)|^{-2}(1+|\lambda|^{2})^{-N_0}e^{|Im \lambda|a}.
\end{eqnarray*}
Integrate the inequality $(a_4)$ between 0 and  $t$ (with respect
$r$) we have
\begin{eqnarray*}
(a_5)\hspace{0.3cm}|F(\lambda,e)|(\int_{0}^{t}e^{-|Im\lambda|r}|\Phi_{\lambda}(r)|\,dr)\leq
C_{N_0} t \, |c(\lambda)|^{-2}(1+|\lambda|^{2})^{-N_0}e^{|Im
\lambda|a}.
\end{eqnarray*}
Consequently
\begin{eqnarray*}
(a_6)\hspace{0.5cm}|F(\lambda,e)|(\frac{1}{t}\int_{0}^{t}e^{-|Im\lambda|r}|\Phi_{\lambda}(r)|\,dr)\leq
C_{N_0} |c(\lambda)|^{-2}(1+|\lambda|^{2})^{-N_0}e^{|Im \lambda|a}
\end{eqnarray*}
for all $\lambda \in l\!\!\!C$. But
\begin{eqnarray*}
1=\lim_{t\to
0}(\frac{1}{t}\int_{0}^{t}e^{-|Im\lambda|r}|\Phi_{\lambda}(r)|\,dr)\,\,\,\,\,\,
for all \lambda\,\,\, \in l\!\!\!C,
\end{eqnarray*}
because
\begin{eqnarray*}
\lim_{t\to
0}(\frac{1}{t}\int_{0}^{t}e^{-|Im\lambda|r}|\Phi_{\lambda}(r)|\,dr)&=&\lim_{t\to
0}(\frac{\Psi(t)}{t}\\ &=&\Psi'(t)|_{t=0}\\
&=&|\Phi_{\lambda}(0)|\\ &=&1
\end{eqnarray*}
with
\begin{eqnarray*}
\Psi(t)=\int_{0}^{t}e^{-|Im\lambda|r}|\Phi_{\lambda}(r)|\,dr.
\end{eqnarray*}
Then $(a_6)$ becomes
\begin{eqnarray*}
|F(\lambda,e)|\leq
C_{N_0}|c(\lambda)|^{-2}(1+|\lambda|^{2})^{-N_0}e^{|Im \lambda|a}.
\end{eqnarray*}
By Di Blasio theorem (see [16]), there exist $f\in
{\mathcal{D}}(NA)^{\#}$ such that $supp f\subset B(e,a)$. In
addition
\begin{eqnarray*}
\frac{F(\lambda,e)}{|c(\lambda)|^{-2}}=\widetilde{f}(\lambda)\,\,\,
for all \,\,\, \lambda \in l\!\!\!C.
\end{eqnarray*}
Whence
\begin{eqnarray*}
\widetilde{f}(\lambda)=\frac{I\!\!P_{\lambda}f(e)}{|c(\lambda)|^{-2}}\frac{4\pi}{c_{m,k}}=\frac{F(\lambda,e)}{|c(\lambda)|^{-2}}.
\end{eqnarray*}
Then
\begin{eqnarray*}
F(\lambda,e)=I\!\!P_{\lambda}f(e)\frac{4\pi}{c_{m,k}}=I\!\!P_{\lambda}f_1(e).
\end{eqnarray*}
Since
\begin{eqnarray*}
\frac{F(\lambda,x)}{F(\lambda,e)}=\frac{I\!\!P_{\lambda}f_1(x)}{I\!\!P_{\lambda}f_1(e)}.
\end{eqnarray*}
We have by the above equality
\begin{eqnarray*}
F(\lambda,x)=I\!\!P_{\lambda}f_1(x),
\end{eqnarray*}
and $f_1 \in {\mathcal{D}}(NA)^{\#}$ with
 $supp f_1\subset B(e,a)$.
\end{proof}
\\
{\bf Conjecture 1.} It will be very interessant to generlize the
theorem 3.1 for the function $f\in
{\mathcal{C}}_{c}^{\infty}(B_a(z))\bigcap {\mathcal{D}}(NA)$. (see
[26] and [27] for the symmetric spaces of non compact of rank
one.)
\\
\begin{Def} Let $T\in {\mathcal{E}}'(NA)$, we define $I\!\!P_{\lambda}T(x)$ as function on $NA$ given by the formula
\begin{eqnarray}
I\!\!P_{\lambda}T(x)&=&\frac{c_{m,k}}{4\pi}|c(\lambda)|^{-2}(T*\Phi_{\lambda})(x)\\
&=&\frac{c_{m,k}}{4\pi}|c(\lambda)|^{-2}<T,\Phi_{\lambda}(d(x,.))>,
\end{eqnarray}
for all $x\in NA$ with $d(x,y)$ denote the distance from $x$ to
$y$.
\end{Def}
\begin{Rem} If $T\in {\mathcal{E}}'(NA)^{\#}$ the above equality becomes for $x=e$
\begin{eqnarray}
I\!\!P_{\lambda}T(e)=\frac{c_{m,k}}{4\pi}|c(\lambda)|^{-2}\widetilde{T}(\lambda)
\end{eqnarray}
Where $\widetilde{T}(\lambda)$ is the spherical Fourier transform
(see [1])
\end{Rem}

\begin{Teo}
Let $T\in {\mathcal{E}}'(NA)^{\#}$ such that suppT$\subset
B_{a}(z)$, then $I\!\!P_{\lambda}T(x)$ satisfies\\ 1) For all
$\lambda \in 1\!\!\!C$, the function $x\to I\!\!P_{\lambda}T(x)$
is a radial function\\ 2)$(\lambda,x)\to I\!\!P_{\lambda}T(x)$ is
a ${\mathcal{C}}^{\infty}$ function on $1\!\!\!C \times NA$\\ 3)
for all $\lambda \in 1\!\!\!C$, we have
${\mathcal{L}}I\!\!P_{\lambda}T(x)=-(\lambda^{2}+\varrho^{2})I\!\!P_{\lambda}T(x)$
(where ${\mathcal{L}}$ is the radial part of the Laplace-Beltrami
operator)\\ 4)  for each fixed x, the function
$I\!\!P_{\lambda}T(x)$ is an entire function divisible by
$|c(\lambda)|^{-2}$ and the quotient is an analytic function\\ 5)
There exists $N_0$ and $C_{N_0}$ such that
\begin{eqnarray*}
|I\!\!P_{\lambda}T(x)|\leq
C_{N_0}|c(\lambda)|^{-2}(1+|\lambda|^{2})^{N_0}e^{|Im
\lambda|(d(x,z)+a)}
\end{eqnarray*}
Conversely, if $x\to F(\lambda,x)$ (for all $\lambda \in
l\!\!\!\!C$) is a radial function and $F(\lambda,x)$ satisfies to
1),2),3),4) and 5) then there exist $T\in {\mathcal{E}}'(NA)^{\#}$
with  suppT$\subset B_{a}(z)$ such that
$I\!\!P_{\lambda}T(x)=F(\lambda,x)$ for all $(\lambda,x) \in
l\!\!\!\!C \times NA)$
\end{Teo}
\begin{proof} The proof of the conditions 1)-- 4) is the same as the proof of  those of theorem 6.2.\\
Now, showing the fifth condition of theorem, recall that for all
$\varphi \in {\mathcal{C}}^{\infty}(NA)^{\#}$, we have for all
$\Delta \in \cup(\underline{NA})$ (see [1]) with,\\ ($\Delta
f$(x)=$\sum \nu_j(x) \frac{d^j}{d\rho^j}f_0(\rho(x)))$ and for
every $g \in {\mathcal{D}}(NA)$ and $x \in NA$ we define the
function $\tau_{x}g$ on $NA$ by the rule
:$\tau_{x}g(y)=g(x^{-1}y)$, $\forall y\in NA$. Since
$\Phi_{\lambda}(x^{-1})=\Phi_{\lambda}(x)$, $\forall x\in NA$ and
supp$\tau_{x}f \subset x supp f$, we obtain
\begin{eqnarray*}
|I\!\!P_{\lambda}T(x)|&=&|\frac{c_{m,k}}{4\pi}|c(\lambda)|^{-2}T*\Phi_{\lambda}(x)|\\
&\leq&
\frac{c_{m,k}}{4\pi}|c(\lambda)|^{-2}|<\tau_{x}T,\Phi_{\lambda}>|,
\end{eqnarray*}
since
\begin{eqnarray*}
|<T,\Phi>|\leq C\sup_{y\in supp T}|D^{r}\Phi(y)|,
\end{eqnarray*}
with C is a constant, then
\begin{eqnarray*}
|I\!\!P_{\lambda}T(x)|&\leq&
c\frac{c_{m,k}}{4\pi}|c(\lambda)|^{-2}\sup_{y\in x^{-1}supp
T}|\frac{d^{m_0}}{d\rho^{m_0}}\Phi_{\lambda}(y)|\\ &=& c'
|c(\lambda)|^{-2}\sup_{y\in
x^{-1}B_{a}(z)}|\frac{d^{m_0}}{d\rho^{m_0}}\Phi_{\lambda}(y)|\\
&=&c' |c(\lambda)|^{-2}\sup_{xy\in
B_{a}(z)}|\frac{d^{m_0}}{d\rho^{m_0}}\Phi_{\lambda}(y)|\\ &=&c'
|c(\lambda)|^{-2}\sup_{d(xy,z)\leq
a}|\frac{d^{m_0}}{d\rho^{m_0}}\Phi_{\lambda}(y)|\\ &=&c'
|c(\lambda)|^{-2}\sup_{d(y,x^{-1}z)\leq
a}|\frac{d^{m_0}}{d\rho^{m_0}}\Phi_{\lambda}(y)|
\end{eqnarray*}
with $c'$ is an absolute constante.\\ since $\{y\in NA
/d(y,x^{-1}z)\leq a\}\subset \{y\in NA /|\rho(y)-d(x,z)|\leq
a\}$,\\ we have
\begin{eqnarray*}
|I\!\!P_{\lambda}T(x)|\leq c'
|{\mathcal{C}}(\lambda)|^{-2}\sup_{\rho(y)\leq
a+d(x,z)}|\frac{d^{m_0}}{d\rho^{m_0}}\Phi_{\lambda}(y)|.
\end{eqnarray*}
Then by Koornwinder lemma (see lemma 3.1), we obtain
\begin{eqnarray*}
|I\!\!P_{\lambda}T(x)|&\leq&
c'_{m_0}|{\mathcal{C}}(\lambda)|^{-2}(1+|\lambda|^2)^{m_0}e^{|Im
\lambda|(a+d(x,z))}
\end{eqnarray*}
with $c'_{m_0}$ is an absolute constante.\\

conversely, assume  $F(\lambda,x)$ satisfies to (1),(2),(3),(4) of
the above theorem, where (4) is verified for some $N_0$. We
construct the distribution $T$  by the rule
\begin{eqnarray*}
<T,\Psi>=\int_{I\!\!R}(\int_{NA}F(\lambda,x)\Psi(x)\,dx)\,d\lambda,
\end{eqnarray*}
for any test function $\Psi $. This is not an absoltely convergent
integral, but we can show that $\int_{NA}F(\lambda,x)\Psi(x)\,dx
\in L^{1}(I\!\!R_\lambda)$.\\ This does not follow directly from
the condition (4) of theorem, but it is easy deduced from it if we
substitute $F(\lambda,x)=
(-\lambda^2)^{-m}({\mathcal{L}}+\rho^2)^{m}F(\lambda,x)$ for all
$m\in I\!\!N$, since
$({\mathcal{L}}+\rho^2)^{m}F(\lambda,x)=(-\lambda^2)^{m}F(\lambda,x)$
and $F(\lambda,x)$ verifie the condition (2). Putting
${\mathcal{L}}_{0}={\mathcal{L}}+\rho^2$, then for all $\Psi \in
{\mathcal{D}}(NA)$ we have
\begin{eqnarray*}
\int_{NA} F(\lambda,x)\Psi(x)\,dx =(-\lambda^2)^{-m}\int_{NA}
({\mathcal{L}}_{0})^{m}F(\lambda,x)\Psi(x)\,dx,
\end{eqnarray*}
an integrating by part means to:
\begin{eqnarray*}
\int_{NA} F(\lambda,x)\Psi(x)\,dx =(-\lambda^2)^{-m}\int_{NA}
F(\lambda,x)({\mathcal{L}}_{0})^{m}\Psi(x)\,dx
\end{eqnarray*}
By the fourth condition of theorem, we have the estimate
\begin{eqnarray*}
|\int_{NA} F(\lambda,x)\Psi(x)\,dx| &\leq&
C_{N_0}(1+|\lambda|^2)^{-m}|I\!\!P_{2r,2l}(\lambda)|(1+|\lambda|^2)^{N_0}e^{a|Im\lambda|}\times\\
&&\times
\int_{supp\Psi}e^{|Im\lambda|d(x,z)}|({\mathcal{L}}_{0})^{m}\Psi(x)|\,dx
\end{eqnarray*}
Using the H$\ddot{o}$lder inequality , the above estimate becomes
\begin{eqnarray*}
|\int_{NA} F(\lambda,x)\Psi(x)\,dx| &\leq&
C'_{N_0,r,l}(1+|\lambda|^2)^{N_0+r_{0}-m}e^{a|Im\lambda|}(\int_{supp\Psi}e^{2|Im\lambda|d(x,z)}\,dx)^{\frac{1}{2}}\\
&&\times
(\int_{NA}|({\mathcal{L}}_{0})^{m}\Psi(x)|^2\,dx)^{\frac{1}{2}}.
\end{eqnarray*}
where $C'_{N_0,r,l}$ is a constant which depend of $N_0$, $r$ and
$l$.\\ Consequently
\begin{eqnarray*}
\int_{I\!\!R}d\lambda|\int_{NA} F(\lambda,x)\Psi(x)\,dx| &\leq&
C''_{N_0,r,l}(\int_{I\!\!R}\frac{d\lambda}{(1+\lambda^2)^{m-N_0-r_{0}}})||({\mathcal{L}}_{0})^{m}\Psi||_{L^{2}(NA)}\\
&<& \infty.
\end{eqnarray*}
Since $m$ is an arbitrary integer, then
\begin{eqnarray*}
\int_{NA} F(\lambda,x)\Psi(x)\,dx \in L^{1}(I\!\!R_{\lambda}).
\end{eqnarray*}
Next we apply a regularization argument. we choose a function
$\widetilde{\theta}_{\epsilon}(\lambda)$ (where
$\theta_{\epsilon}(x)$ is the regularised function (see [1]), and
$\widetilde{\theta}_{\epsilon}(\lambda)$ is the spherical Fourier
transform. From [17] theorem 3.5, we have
\begin{eqnarray*}
|\widetilde{\theta}_{\epsilon}(\lambda)|\leq
C_{\epsilon}(1+|\lambda|)^{-n_0}e^{|Im\lambda|\epsilon}.
\end{eqnarray*}
The function $\widetilde{\theta}_{\epsilon}(\lambda)F(\lambda,x)$
verifie the conditions of theorem 3.1 in [1]. Then, there exists a
function $F_{\epsilon}\in {\mathcal{D}}(NA)^{\#}$ such that:
$suppF_{\epsilon}\subset B_{a+\epsilon}(z)$ and
$I\!\!P_{\lambda}F_{\epsilon}=\widetilde{\theta}_{\epsilon}(\lambda)F(\lambda,x)$
and $\int_{I\!\!R}
\widetilde{\theta}_{\epsilon}(\lambda)F(\lambda,x)\Psi(x)\,d\lambda=
F_{\epsilon}(x)$ as $\widetilde{\theta}_{\epsilon} \to 1$ as
$\epsilon \to 0$ (see [1]), we have then
\begin{eqnarray*}
\int_{NA}F_{\epsilon}(x)\Psi(x)\,dx=\int_{NA}(\int_{I\!\!R}\widetilde{\theta}_{\epsilon}(\lambda)F(\lambda,x)\Psi(x))\,dx
d\lambda,
\end{eqnarray*}
when $\epsilon \to 0$ the above equality becomes
\begin{eqnarray*}
\lim_{\epsilon \to
0}\int_{NA}F_{\epsilon}(x)\Psi(x)\,dx&=&\int_{NA}(\int_{I\!\!R}F(\lambda,x)\Psi(x)\,dx)\,d\lambda\\
&=& <T,\Psi>.
\end{eqnarray*}
Whence , when $\epsilon \to 0$, we have also
\begin{eqnarray*}
<F_{\epsilon},\Psi>\to <T,\Psi> \hspace{1cm}\forall \Psi \in
{\mathcal{D}}(NA).
\end{eqnarray*}
As $suppF_{\epsilon}\subset B_{a+\epsilon}(z)$ and
$I\!\!P_{\lambda}F_{\epsilon}=\widetilde{\theta}_{\epsilon}(\lambda)F(\lambda,x)$,
we obtain \\ $I\!\!P_{\lambda}T(x)=F(\lambda,x)$ and $suppT\subset
B_{a}(z)$, and this completes the proof.
\end{proof}
 \begin{Rem}1):  If $T=\delta^{n_0}$, with $\delta^{n_0}$ the derivation of  the Dirac measure $\delta_{e}$, the spectral projection operator of $T=\delta ^{n_0}$ becomes
\begin{eqnarray*}
I\!\!P_{\lambda}T(x)&=&\frac{c_{m,k}}{4\pi}|c(\lambda)|^{-2}\delta^{n_0}*\Phi_{\lambda}(x)\\
&=&\frac{c_{m,k}}{4\pi}|c(\lambda)|^{-2}\frac{d^{n_0}}{d\rho^{n_0}}\Phi_{\lambda}(x).
\end{eqnarray*}
According to the Koornwinder lemma (see lemma 3.1), the above
equality becomes
\begin{eqnarray*}
|I\!\!P_{\lambda}T(x)|\leq
\frac{c_{m,k}}{4\pi}|c(\lambda)|^{-2}(1+|\lambda|^{2})^{n_0}e^{|Im\lambda|\rho(x)},
\end{eqnarray*}
we find then the result of [1].\\ \hspace{1cm}2): For $x=z=e$ and
$T$ a radial compactly supported distribution in $\{e\}$, the
spectral projection becomes also (when $dim NA$ is odd) (see [1])
\begin{eqnarray*}
I\!\!P_{\lambda}T(x)&=&\frac{c_{m,k}}{4\pi}|c(\lambda)|^{-2}\widetilde{T}(\lambda)\\
&\leq &c_{n_0}|c(\lambda)|^{-2}(1+|\lambda|)^{n_0}.
\end{eqnarray*}
\end{Rem}
{\bf Conjecture 2.} do we have a generalisation of theorem 3.2 for
the distributions which is not necessary radials?

\section{Characterization of the range of ${\mathcal{S}}^{p}(NA)^{\#}$ by spectral projection operator}
Let $\Omega$ be a left invariant differential operator on $NA$ of
order $l$, defined as follows
\begin{eqnarray*}
\Omega f(x)=\sum_{j=1}^{j=l} \mu_{j}(x)
\frac{d^j}{d\rho^j}f_0(\rho(x)))\hspace{0.5cm} \forall f\in
{\mathcal{C}}^{\infty}(NA).
\end{eqnarray*}
where $\mu_{j}$ (j=1,2,..,l) the ${\mathcal{C}}^{\infty}$
functions on $NA$. From lemma 2.3 in [17] p. 28, there exists a
constant $c$ depending only on $\Omega$ such that
\begin{eqnarray*}
\sup_{\rho(x)\geq 0}|\mu_{j}(x)|\leq c.
\end{eqnarray*}
\\
For $0<p\leq 2$ denote by, (see [17]),
${\mathcal{S}}^{p}(NA)^{\#}$ the space of radial and
${\mathcal{C}}^{\infty}$ functions $f$ on $NA$ such that
\begin{eqnarray}
v_{p}(f,\Omega ,h)&=&\sup_{x\in
NA}e^{\frac{Q}{p}\rho(x)}(1+\rho(x))^{h}|\Omega f(x)|\\ &<&\infty
\nonumber
\end{eqnarray}
for all positive integers $h$ and all left invariant differential
operators $\Omega$ on $NA$.\\ We can define the space
${\mathcal{S}}^{p}(NA)^{\#}$ in a different way (see [3]) instead
of (4.1) we use the condition
\begin{eqnarray*}
\sup_{\rho \geq
0}e^{\frac{Q}{p}\rho(x)}(1+\rho(x))^{h}|\frac{d^{l}}{d\rho^{l}}f_{0}(\rho)|<\infty
\hspace{1cm}\forall f\in {\mathcal{C}}^{\infty}
\end{eqnarray*}
For $\epsilon >0$ define $\Omega _{\epsilon}=\{s\in l\!\!\!C:
|Res|<\epsilon \frac{Q}{2}\}$. Denote also by
${\mathcal{H}}(\Omega _{\epsilon})$ the space of
${\mathcal{C}}^{\infty}$ function $\phi$ on $\Omega _{\epsilon}$
such that $\phi(s)=\phi(-s)$ for all s$\in \Omega _{\epsilon}$ and
such that
\begin{eqnarray*}
{\mathcal{V}}_{\epsilon}(\phi,l,h)&=&\sup_{|Res|<\epsilon
\frac{Q}{2}}(1+|s|)^{h}|\frac{d^{l}}{ds^{l}}\phi(s,x)|\\ &<&\infty
\end{eqnarray*}
for all positive integers h and l. consider on
${\mathcal{H}}(\Omega _{\epsilon})$ the topology defined by the
semi-normes ${\mathcal{V}}_{\epsilon}(\phi,l,h)$ (see [17] p. 34).
\begin{Lem} (see [17], p. 34) Let $0<p\leq 2$ and $\epsilon=\frac{2}{p}-1$. Then the spherical transform $f\to \widetilde {f}$ is a topological isomorphims from ${\mathcal{S}}^{p}(NA)^{\#}$ onto ${\mathcal{H}}(\Omega _{\epsilon})$
\end{Lem}
\begin{Pro} The function $f$ is radial if and only if $I\!\!P_{\lambda}f$ is radial
\end{Pro}
\begin{proof}
If $f$ is radial, from the inversion formula of the spherical
transform the spectral projection is
\begin{eqnarray}
I\!\!P_{\lambda}f(x)=\frac{c_{m,k}}{4\pi}|c(\lambda)|^{-2}
\widetilde{f}(\lambda)\Phi_{\lambda}(x),
\end{eqnarray}
as $\Phi_{\lambda}$ is radial we have then that
$I\!\!P_{\lambda}f$ is radial, and the reverse follow from the
rule
\begin{eqnarray*}
f(x)=\int_{-\infty}^{+\infty}I\!\!P_{\lambda}f(x)\,d\lambda.
\end{eqnarray*}
\end{proof}
\\
It follows that from (4.2) that
\begin{eqnarray*}
I\!\!Pf(e)=\frac{c_{m,k}}{4\pi}|c(\lambda)|^{-2}\widetilde{f}(\lambda).
\end{eqnarray*}
One may characterize the ${\mathcal{S}}^{p}(NA)^{\#}$- range via
the spectral projection $I\!\!Pf(x)$ for any $x\in NA$ and
$\lambda \in \Omega _{\epsilon}$. Let now define
${\mathcal{H}}(\Omega _{\epsilon}\times NA)$ the space of
${\mathcal{C}}^{\infty}$ function F on $\Omega _{\epsilon}\times
NA$ such that $F(-\lambda,x)=F(\lambda,x)$ and
$F(\lambda,x^{-1})=F(\lambda,x)$ for all $\lambda \in \Omega
_{\epsilon}$ and  $x\in NA$ and
${\mathcal{L}}_rF(\lambda,.)=-(\frac{Q^{2}}{4}+\lambda^{2})F(\lambda,.)$,
and for any left invariant differential operator $D$ on $NA$ of
order $l$, such that
\begin{eqnarray*}
p_{(\epsilon,N,D)}(F)&=&\sup_{|Re\lambda|<\epsilon
\frac{Q}{2},x\in
NA}(1+|\lambda|)^{N}e^{-|Im\lambda|d(x,e)}|DF(\lambda,x)|\\
&<&\infty
\end{eqnarray*}
for all positive integers $N$ .
\begin{Teo} Let $0<p\leq 2$ and $\epsilon=\frac{2}{p}-1$. Then the spectral projection transform $f\to I\!\!Pf$ is a topological isomorphims from ${\mathcal{S}}^{p}(NA)^{\#}$ onto ${\mathcal{H}}(\Omega _{\epsilon}\times NA)$
\end{Teo}
The theorem 4.1 deduce from the following Theorem after having
using the closed graph theorem.

\begin{Teo}Let $0<p\leq 2$ and $\epsilon=\frac{2}{p}-1$. There exists $f\in {\mathcal{S}}^{p}(NA)^{\#}$ such that $I\!\!P_{\lambda}f(x)=I\!\!F(\lambda,x)$ if and only if\\
{\bf 1)} a) For each $x\in NA$ we have
$F(-\lambda,x)=F(\lambda,x)$ for all $\lambda \in l\!\!\!C$\\

b) For each fixed $\lambda$, we have
$F(\lambda,x)=F(\lambda,x^{-1})$ for all $x\in NA$\\

c) $(\lambda,x)\to F(\lambda,x)$ is radial and
${\mathcal{C}}^{\infty}$ function on $\Omega _{\epsilon}\times
NA$\\
\\
{\bf2)} for each $\lambda$, we have
${\mathcal{L}}_rF(\lambda,x)=-(\lambda^{2}
+\frac{Q^2}{4})F(\lambda,x)$\\
\\
{\bf3)} for each $N $and each left differential operator $D$ of
order $l$ on $NA$, there exists $c_{N,D}$ such that
\begin{eqnarray*}
|DI\!\!F(\lambda,x)|\leq
c_{\epsilon,N,D}|c(\lambda)|^{-2}(1+|\lambda|)^{-N+l}e^{|Im\lambda|d(e,x)}\hspace{0.5cm}
if {\hspace{0.5cm}}|Re\lambda|<\epsilon \frac{Q}{2}.
\end{eqnarray*}
\end{Teo}
\begin{proof} Assume that $f\in {\mathcal{S}}^{p}(NA)^{\#}$ ($0<p\leq 2$), since $f$ is radial, the spectral projection becomes
\begin{eqnarray}
I\!\!P_{\lambda}f(x)=\frac{c_{m,k}}{4\pi}|c(\lambda)|^{-2}
\widetilde{f}(\lambda)\Phi_{\lambda}(x).
\end{eqnarray}
From the above lemma, we have that $\widetilde{f}(\lambda)$ is
${\mathcal{C}}^{\infty}$ on $\Omega_{\epsilon}$ and since
$\Phi_{\lambda}(x)$ is ${\mathcal{C}}^{\infty}$ function on
$\Omega_{\epsilon}\times NA$, then 1) and 2) of the theorem
follows immediatly. Now showing the third condition. From the
lemma 4.1, there exists a constant $C$ such that
\begin{eqnarray}
 |\frac{d^{k}}{d\lambda^{k}}\widetilde{f}(\lambda)|\leq C(1+|\lambda|)^{-N}\hspace{0,5cm}if \hspace{0,3cm}|Re\lambda|< \epsilon \frac{Q}{2}
\end{eqnarray}
for all positive integers $N$ and $k$, where $C$ depend of
$N,\epsilon,k$ but not of $\lambda$. According to the Koornwinder
lemma, there exists a constant $C'$ such that
\begin{eqnarray}
|\frac{d^{n}}{dr^{n}}\Phi_{\lambda}(r)|\leq
C'(1+|\lambda|)^{n}e^{r|Im\lambda|}(1+r)e^{-r\frac{m+2k}{2}},
\end{eqnarray}
we have $(1+r)e^{-r\frac{m+2k}{2}} \to 0$ as $r\to \infty$. Then
the inequality 4.5 becomes
\begin{eqnarray}
 |\frac{d^{n}}{dr^{n}}\Phi_{\lambda}(r)|\leq C''(1+|\lambda|)^{n}e^{r|Im\lambda|},
\end{eqnarray}
with $C''$ is an other constant. Using  the formulas 4.5 and 4.6
to obtain that
\begin{eqnarray*}
|\Omega
I\!\!P_{\lambda}f(x)|&=&|\sum_{j=1}^{l}\mu_{j}(x)\frac{d^{j}}{d\rho^{j}}I\!\!P_{\lambda}f(\rho)|\\
&=&\frac{c_{m,k}}{4\pi}|c(\lambda)|^{-2}
|\sum_{j=1}^{l}\mu_{j}(x)\frac{d^{j}}{d\rho^{j}}\Phi_{\lambda}(x)||\widetilde{f}(\lambda)|.
\end{eqnarray*}
Then from the formulas (4.4) and (4.6), there exists a constant
$c'_{\epsilon,N}$ such that
\begin{eqnarray*}
|\Omega I\!\!P_{\lambda}f(x)|&=&|\Omega I\!\!P_{\lambda}f(\rho)|\\
&\leq&
c'_{\epsilon,N}\frac{c_{m,k}}{4\pi}|c(\lambda)|^{-2}(1+|\lambda|)^{-N+l}e^{|Im\lambda|\rho}\hspace{0.5cm}
if |Re\lambda|<\epsilon \frac{Q}{2}.
\end{eqnarray*}
for all positive integers $N$, with $l$ is the order of $\Omega
$.\\ Conversely, assume that there exists a radial function
$I\!\!F(\lambda,x)$ that satisfies to 1), 2) and 3) of the
theorem. putting
\begin{eqnarray}
\Psi(\lambda)=\frac{4\pi}{c_{m,k}}\frac{I\!\!F(\lambda,x)}{\phi_{\lambda}(x)}|c(\lambda)|^{-2},
\end{eqnarray}
since
\begin{eqnarray*}
{\mathcal{L}}_r(\frac{I\!\!F(\lambda,x)}{I\!\!F(\lambda,e)})&=&-(\lambda^{2}+\frac{Q^2}{4})\frac{I\!\!F(\lambda,x)}{I\!\!F(\lambda,e)},\\
\frac{I\!\!F(\lambda,e)}{I\!\!F(\lambda,e)}&=&1
\end{eqnarray*}
and $I\!\!F(\lambda,x)$ is radial, we have
\begin{eqnarray}
I\!\!F(\lambda,x)=I\!\!F(\lambda,e)\phi_{\lambda}(x).
\end{eqnarray}
From the fourth condition of theorem, we have for all $N\in
I\!\!N$
\begin{eqnarray*}
|I\!\!F(\lambda,x)|\leq
c'_{\epsilon,N}(1+|\lambda|)^{-N}e^{|Im\lambda|d(e,x)},
\end{eqnarray*}
according to the formula (4.7), the above formula becomes
\begin{eqnarray*}
|c(\lambda)|^{2}|\phi_{\lambda}(x)||\Psi(\lambda)|\leq
c'_{\epsilon,N}(1+|\lambda|)^{-N+l}e^{|Im\lambda|d(e,x)}.
\end{eqnarray*}
Since $\sup_{\rho \geq 0}\Phi_{\lambda}(\rho)\leq c'e^{\rho
|Im\lambda|}$ (see formula (4.6)) and that there exists a constant
$C_{1}$ and a constant b (see formula 7.4 in [6], p. 25 and [17]
p. 37) such that
\begin{eqnarray*}
|c(\lambda)|^{-2}\leq
C_{1}(1+|\lambda|)^{b}\hspace{0.5cm}|Re\lambda|\leq
\epsilon\frac{Q}{2}
\end{eqnarray*}
we obtain then , for any $N$, that
\begin{eqnarray*}
|\Psi(\lambda)|\leq C''_{\epsilon,N}(1+|\lambda|)^{-N+l+b}.
\end{eqnarray*}
Where $C''_{\epsilon,N}$ is constant which depend only of $N$,
$\Omega$ and $\epsilon$. Since $N$ is an arbitrary positive
integer and according to the lemma 4.1, there exists a function
$f\in {\mathcal{S}}^{p}(NA)^{\#}$ with $0<p\leq 2$ such that
$\widetilde{f}(\lambda)=\Psi(\lambda)$, then
\begin{eqnarray*}
\widetilde{f}(\lambda)=\frac{4\pi}{c_{m,k}}\frac{I\!\!F(\lambda,x)}{\phi_{\lambda}(x)}|c(\lambda)|^{-2},
\end{eqnarray*}
consequentely,
\begin{eqnarray*}
I\!\!F(\lambda,x)=\frac{c_{m,k}}{4\pi}|c(\lambda)|^{-2}\phi_{\lambda}(x)\widetilde{f}(\lambda),
\end{eqnarray*}
which is equal to $I\!\!P_{\lambda}f(x)$ since $I\!\!F$ is radial.
\end{proof}

\section{$L^{2}$-Estimation for spectral projection operator}
The aim of this section is to do the $L^{2}$-estimation for
spectral projection. Recall that the Plancherel's formula (see
[17]) for Fourier spherical transform of $f\in L^{2}(NA)^{\#}$ is
\begin{eqnarray}
||f||_2^2=\frac{c_{m,k}}{2\pi}\int_{0}^{\infty}|\widetilde{f}(\lambda)|^{2}|c(\lambda)|^{-2}\,d\lambda.
\end{eqnarray}
\begin{Teo} For $x\in NA$ and $f\in L^{2}(NA)$, the following inequality holds
\begin{eqnarray}
\int_{0}^{\infty}|I\!\!P_{\lambda}f(x)|^{2}|c(\lambda)|^{2}\,d\lambda
\leq \frac{c_{m,k}}{8\pi}||f||_{L^2(NA)}^2
\end{eqnarray}
\end{Teo}
\begin{proof}  Let $f\in L^{2}(NA)$ and $f_{x}(y)=M(\tau_{x^{-1}}f)(y)$ the averaging function of the translated function $\tau_{x^{-1}}f$ (see [24]), we remark that if $f\in L^{2}(NA)$ then $f_{x}\in L^{2}(NA)^{\#}$ for any $x\in NA$ (since $NA$ is endowed with a left Haar measure and $||Mf||_{2}\leq ||f||_{2}$ (see [14])). From this, the formula 5.1 becomes
\begin{eqnarray}
||f_{x}||_2^2=\frac{c_{m,k}}{2\pi}\int_{0}^{\infty}|\widetilde{f_{x}}(\lambda)|^{2}|c(\lambda)|^{-2}\,d\lambda,
\end{eqnarray}
and since
\begin{eqnarray*}
I\!\!P_{\lambda}f(x)&=&\frac{c_{m,k}}{4\pi}|c(\lambda)|^{-2}(f*\Phi_{\lambda})(x)\\
&=&\frac{c_{m,k}}{4\pi}|c(\lambda)|^{-2}\widetilde{f_{x}}(\lambda).
\end{eqnarray*}
Using the above equality in the formula (5.3) to obtain that, for
every $x\in NA$
\begin{eqnarray*}
||f_{x}||_2^2&=&||M(\tau_{x^{-1}}f)||_2^2\\
&=&\frac{8\pi}{c_{m,k}}\int_{0}^{\infty}|I\!\!P_{\lambda}f(x)|^{2}|c(\lambda)|^{2}\,d\lambda.
\end{eqnarray*}
From the Proposition 1.3 in [14], the following properties hold
for every ,$f\in L^{p}(NA)$, with $1\leq p\leq \infty$
\begin{eqnarray*}
||Mf||_{p}\leq ||f||_{p} .
\end{eqnarray*}
Then, we will have the following formula for any $x\in NA$
\begin{eqnarray}
\int_{0}^{\infty}|I\!\!P_{\lambda}f(x)|^{2}|c(\lambda)|^{2}\,d\lambda
&\leq& \frac{c_{m,k}}{8\pi}||\tau_{x^{-1}}f||_2^2\\
&=&\frac{c_{m,k}}{8\pi}||f||_2^2,
\end{eqnarray}
\end{proof}
\begin{Teo} Let $K$ be a compact set of $NA$ and $x$ an element of a compact $K$, we assume that $f\in L^{2}(NA)$, then we have the following estimate
\begin{eqnarray*}
\int_{-\infty}^{\infty}|c(\lambda)|^{2}|I\!\!P_{\lambda}f(x)|^{2}\,d\lambda
\leq
\frac{2^{m}\pi^{\frac{m+k}{2}}}{\Gamma(\frac{m+k}{2})}\frac{c_{m,k}}{4\pi}c(K)||f||_{2}^{2},
\end{eqnarray*}
with c(K)a constant which depend only of K.
\end{Teo}
\begin{proof} Let $f$ be an element of $L^{2}(NA)$ and $x$ an element of a compact $K$ in $NA$, if we use the spectral projector as (see formula (1.6))
\begin{eqnarray}
I\!\!P_{\lambda}f(x)=\frac{c_{m,k}}{4\pi}|c(\lambda)|^{-2}\int_{N}{\mathcal{P}}_{-\lambda}(x,n)\widehat{f}(\lambda,n)\,dn.
\end{eqnarray}
we will have the same result as the proposition 5.1.\\ Using the
H$\ddot{o}$lder inequality in the formula (5.5) to obtain that
\begin{eqnarray*}
|I\!\!P_{\lambda}f(x)|^{2}&=&(\frac{c_{m,k}}{4\pi})^{2}|c(\lambda)|^{-4}(\int_{N}{\mathcal{P}}_{-\lambda}(x,n)\widehat{f}(\lambda,n)\,dn)^{2}\\
&\leq&
(\frac{c_{m,k}}{4\pi})^{2}|c(\lambda)|^{-4}(\int_{N}|{\mathcal{P}}_{-\lambda}(x,n)|^2\,dn)\times\\
&&(\int_{N}|\widehat{f}(\lambda,n)|^2\,dn)\\ &\leq&
(\frac{c_{m,k}}{4\pi}|c(\lambda)|^{-2}\int_{N}|{\mathcal{P}}_{-\lambda}(x,n)|^2\,dn)\times\\
&&(\frac{c_{m,k}}{4\pi}\int_{N}|\widehat{f}(\lambda,n)|^{2}|c(\lambda)|^{-2}\,dn)
\end{eqnarray*}
then
\begin{eqnarray}
|c(\lambda)|^{2}|I\!\!P_{\lambda}f(x)|^{2}&\leq&(\frac{c_{m,k}}{4\pi}\int_{N}|{\mathcal{P}}_{-\lambda}(x,n)|^2\,dn)\times\\
&&(\frac{c_{m,k}}{4\pi}\int_{N}|\widehat{f}(\lambda,n)|^{2}|c(\lambda)|^{-2}\,dn).\nonumber
\end{eqnarray}
According to the Plancherel formula for Fourier-Helgason transform
(see [7]) and since ${\mathcal{P}}_{-\lambda}(x,n)\in L^{2}(N)$
(see [24], p. 44), we obtain that, for all compact $K$ of $NA$,
there exists a constant $c(K)$ such that
\begin{eqnarray*}
|{\mathcal{P}}_{-\lambda}(x,n)|&\leq& c(K)e^{\rho (t\circ
\sigma)n}\\ &&\in L^{2}(N),
\end{eqnarray*}
the formula (5.6) becomes
\begin{eqnarray*}
\int_{-\infty}^{\infty}|c(\lambda)|^{2}|I\!\!P_{\lambda}f(x)|^{2}\,d\lambda
\leq \frac{c_{m,k}}{4\pi}c(K)||f||_{2}(\int_{N}e^{2\rho (t\circ
\sigma)n}\,dn),
\end{eqnarray*}
since $\int_{N}e^{2\rho (t\circ \sigma)n}\,dn=
2^{-k}|{\mathcal{S}}^{m+k-1}|=2^{-k}\frac{2^{n-1}\pi^{\frac{n-1}{2}}}{\Gamma(\frac{n-1}{2})}$
(with $n=m+k+1$)(see [24], p. 44), then
\begin{eqnarray*}
\int_{-\infty}^{\infty}|c(\lambda)|^{2}|I\!\!P_{\lambda}f(x)|^{2}\,d\lambda
\leq
2^{-k}\frac{2^{n-1}\pi^{\frac{n-1}{2}}}{\Gamma(\frac{n-1}{2})}\frac{c_{m,k}}{4\pi}c(K)||f||_{L^{2}(NA)}.
\end{eqnarray*}
\end{proof}
\section{Description of the eigenspace of the invariant Laplacian $\Delta$ on $D$}
For every complex number $\lambda \in l\!\!\!C$, let
$E_{\lambda}(D)$ be the space of all eigenfunctions of
$\Delta_{D}$ in $D$ with eigenvalue $-(\lambda^2 +1)$. since the
operator $\Delta_{D}$ is elliptic in $D$, the elements of
$E_{\lambda}(D)$  are ${\mathcal{C}}^{\infty}$-functions on $D$
i,e.,
\begin{eqnarray}
E_{\lambda}(D)=\{F\in {\mathcal{C}}^{\infty}(D); \Delta_{D} F =
-(\lambda^2 +1)F\}.
\end{eqnarray}
Now, let ${\mathcal{H}}_{k}$ denotes the space of restrictions to
$S^{1}=\partial D$ of harmonic polynomials $z^k$ and
$\overline{z}^k$ which are homogeneous of degree $k$ in $z$. Then,
it is well known that ${\mathcal{H}}_{k}$ is $SO(2)$-irreductible
and we have $L^2 (S^{1})=\oplus_{k\in 1\!\!Z}{\mathcal{H}}_{k}$.
\begin{Pro}(see [9])
A function $F$ is in the eigenspace $E_{\lambda}(D)$, if and only
if $F$ can be expanded in ${\mathcal{C}}^{\infty}(D)$ as
\begin{eqnarray}
&&\\ F(z)&=&\sum_{k\in l\!\!Z}e^{ik\theta }a_{k}(\lambda) (\tanh
r)^{|k|}\,
_{2}F_{1}(\frac{1+i\lambda}{2},\frac{1-i\lambda}{2};1+|k|;-\sinh^{2}(r))
\nonumber,
\end{eqnarray}
where $a_{k}(\lambda)$ is a constant which depend only of $k$ and
$\lambda$.
\end{Pro}
{\bf Proof}. See [9] for the proof of this proposition.\\
\\
The generalized spherical function is given by
\begin{eqnarray}
&&\\ \Phi_{\lambda,k}(\tanh
r)&=&\int_{S^1}{\mathcal{P}}_{\lambda}(z,e^{i\theta})e^{i k
\theta}\,d\sigma(\theta)\nonumber \\ &=&(1-(\tanh
r)^{2})^{\frac{1+i\lambda}{2}}|\tanh r|^{|k|}\frac{\Gamma
(|k|+\frac{1+i\lambda}{2})}{\Gamma (\frac{1+i\lambda}{2})|k|!}\,
_{2}F_{1}(\frac{1+i\lambda}{2},|k|\nonumber\\
&&+\frac{1+i\lambda}{2};1+|k|;(\tanh(r))^{2})\nonumber \\
&=&|\tanh r|^{|k|}\frac{\Gamma (|k|+\frac{1+i\lambda}{2})}{\Gamma
(\frac{1+i\lambda}{2})|k|!}\,
_{2}F_{1}(\frac{1+i\lambda}{2},\frac{1-i\lambda}{2};1+|k|;-\sinh^{2}(r))\nonumber.
\end{eqnarray}
Now, let $X_{k,\lambda}$ denote the one-dimensional space spanned
by the function
\begin{eqnarray}
\Phi_{\lambda,k}(\tanh r)e^{i k \theta}.
\end{eqnarray}
We note that
\begin{eqnarray}
E_{\lambda}= \oplus_{k\in l\!\!\!Z}X_{k}.
\end{eqnarray}
Let $n$ be choosen fixed in $l\!\!\!Z$ so that we have
\begin{eqnarray}
E_{\lambda}= \oplus_{k\geq |n|}X_{k} \oplus \oplus_{k\leq
|n|}X_{k},
\end{eqnarray}
Let \begin{eqnarray} E'_{n}= \oplus_{k\geq |n|}X_{k}\,\,\,\,and
\,\,\,\, E''_{n}=\oplus_{k\leq |n|}X_{k},
\end{eqnarray}
We denote by
\begin{eqnarray}
\widetilde{E}_{\lambda}(D)=\{F\in {\mathcal{C}}^{\infty}(D);
(\Delta_{D} +\lambda^2 +1)^{2}F= 0\}.
\end{eqnarray}
It is easy to see that $E_{\lambda}(D)\subset
\widetilde{E}_{\lambda}(D)$.\\
\\ We need to describe the functions in $\widetilde{E}_{\lambda}(D)$ that are not in $E_{\lambda}(D)$. We observe that: If $g_{\lambda}\in E_{\lambda}(D)$ then $\frac{d}{d\lambda}g_{\lambda}\in \widetilde{E}_{\lambda}(D)$. In fact, it follows from
\begin{eqnarray}
0=\frac{d}{d\lambda}(\Delta_{D} +\lambda^2
+1)g_{\lambda}=(\Delta_{D} +\lambda^2
+1)\frac{d}{d\lambda}g_{\lambda}+ 2\lambda g_{\lambda}.
\end{eqnarray}
We say that $f$ is $SO(2)$-finite if $f(\tanh r e^{i\theta})$ can
be expanded into a finite spherical series expansion with respect
to $e^{i \theta}$ to get $$f(\tanh r e^{i\theta})=\sum_{|k|\leq
m_0}f_{\lambda,k}(\tanh r) e^{ik \theta}$$ is
${\mathcal{C}}^{\infty}(I\!\!R^+\times S^1)$, for ceratain $m_0
\in l\!\!Z^{+}$
\begin{Lem}
The following assertion is equivalent\\ i)\hspace{0.5cm} $f(z)$ is
$SO(2)$-finite \\ ii)$\widehat{f}(\lambda,w)$ is $SO(2)$-finite \\
iii)$I\!\!P_{\lambda}f(z)$ is $SO(2)$-finite .
\end{Lem}
{\bf Proof.} Let $f(z)$ be $SO(2)$-finite, then
\begin{eqnarray*}
\widehat{f}(\lambda,w)&=&\int_{D}{\mathcal{P}}_{-\lambda}(z,w)f(z)\,d\mu(z)\\
&=&\sum_{|k|\leq
m_0}\int_{D}{\mathcal{P}}_{-\lambda}(z,w)f_{\lambda,k}(\tanh r)
e^{ik \theta}\,d\mu(z)\,\,\,\,with \,\,\,z=\tanh re^{i \theta} \\
&=&\sum_{|k|\leq
m_0}\int_{D}(\frac{1-|z|^{2}}{1-z.\overline{w}})^{\frac{1-i\lambda}{2}}f_{\lambda,k}(\tanh
r) e^{ik \theta}\,d\mu(z)\\ &=&\sum_{|k|\leq
m_0}(\int_{0}^{1}(1-r^2)^{\frac{1-i\lambda}{2}-2}r
f_{\lambda,k}(\tanh
r)\,dr)(\int_{S^1}(\frac{1-r^2}{|1-r.w|^2})^{\frac{1-i\lambda}{2}}e^{ik
\theta}\,d\sigma(\theta))\\ &=&\sum_{|k|\leq m_0}a_{\lambda,k}w^k
\end{eqnarray*}
where $w^k=e^{ik \varphi}$ and $a_{\lambda,k}= \frac{\Gamma
(|k|+\frac{1+i\lambda}{2})}{\Gamma
(\frac{1+i\lambda}{2})|k|!}\int_{0}^{1}(1-r^{2})^{\frac{1-i\lambda}{2}-2}f_{\lambda,k}(\tanh
r)\,
_{2}F_{1}(\frac{1+i\lambda}{2},|k|+\frac{1+i\lambda}{2};1+|k|;(\tanh(r))^{2})\,dr$,
then the assertion i) implique ii).\\ The assertion ii) $\to$ iii)
is obtained from the formula of $I\!\!P_{\lambda}f(z)$ and the
same proof as the above. The assertion iii)$\to$ i) is deduced
from the inversion formula in the formula
$I\!\!P_{\lambda}f(z)$.\\
\\
Let $B_{R}(z_{0})$ design the ball of radius $R$ and center $z_0$.
\section{Spectral projection operator on ${\mathcal{C}}_{com}^{\infty}(D)$ associated to the Laplacian $\Delta$}
In this section, we define the spectral projection oprator
$I\!\!P_{\lambda}f(z)$ on ${\mathcal{C}}_{com}^{\infty}(D)$
associated to the Laplacian $\Delta_{D}$, we give an expression
exhibits $I\!\!P_{\lambda}f(z)$ as a meromorphic function of
$\lambda$, making appear the poles and zeros of
$I\!\!P_{\lambda}f(z)$. We begin by givining the necessary
condition.
\begin{Teo}
Suppose $f$ is ${\mathcal{C}}^{\infty}$ with support in
$\overline{B_{R}(z_{0})}$, $f$ is $SO(2)$-finite then\\
\\
1)\hspace{0.5cm} $I\!\!P_{\lambda}f(z)$ is $C^{\infty}$ function
on $(l\!\!\!C-il\!\!\!Z)\times D$\\
\\
2)$\hspace{0.5cm}$for each fixed $\lambda \in l\!\!\!C-il\!\!\!Z$,
we have  $\Delta_{D}
I\!\!P_{\lambda}f(z)=-(\lambda^{2}+1)I\!\!P_{\lambda}f(z)$\\
\\
3)$\hspace{0.5cm}$for each fixed $z$, $I\!\!P_{\lambda}f(z)$ is an
even function meromorphic function of $\lambda$ with at worst
simple poles at $\lambda_{k}=\frac{+}{}i(2k+1)$  for $k\geq |n|$
where $n$ is choosed  fixed in $l\!\!\!Z$), and
\begin{eqnarray}
\sum_{k\in
l\!\!\!Z}Res_{\lambda=\lambda_{k}}I\!\!P_{\lambda}f(z)=0
\end{eqnarray}
4)$\hspace{0.5cm}$for every $N$ there exists $c_N$ such that
\begin{eqnarray}
|I\!\!P_{\lambda}f(z)|\leq c_{N}(1+|\lambda|
)^{-N}e^{(R+d(z,z_{0}))|Im \lambda|}
\end{eqnarray}
5)\hspace{0.5cm}$I\!\!P_{\lambda}f(z)$ has a simple zeros at
points $\lambda_l=\frac{+}{} i 2l$ ($l\in l\!\!Z^{*})$ and a
double zero at $\lambda=0$ and  satisfies\\
\\
$\bullet$ $z\to Res_{\lambda=\lambda_{k}}I\!\!P_{\lambda}f(z)\in
E'_{n}$\\
\\
$\bullet$ $(I\!\!P_{\lambda}f(z)-(\lambda
-\lambda_{k})^{-1}Res_{\lambda=\lambda_{k}})\vert_{\lambda=\lambda_{k}}
\in \widetilde{E''}_{n}$\\
\\
$\bullet$ $I\!\!P_{\lambda}f(z)\vert_{\lambda=0}=0$ and
$\frac{I\!\!P_{\lambda}f(z)}{\Gamma(|k|+\frac{1+i\lambda}{2})\Gamma(|k|+\frac{1-i\lambda}{2})}$
has even entire expansion.
\end{Teo}
In order to do the proof of this theorem, we need some preparatory
results
\begin{Pro}
Let $f$ be an element  of ${\mathcal{C}}_{com}^{\infty}(D)$, then
\begin{eqnarray}
I\!\!P_{\lambda}f(z)=(f*\varphi_{\lambda
})(z)=\int_{D}\varphi_{\lambda}(d(z,z'))f(z')\,dz',
\end{eqnarray}
where
\begin{eqnarray}
\varphi_{\lambda}(\tanh r)&=& \frac {(2\pi
)^{(\frac{1}{4})}}{4\pi^{2}}\lambda \tanh(\frac{\pi
\lambda}{2})\Phi_{\lambda }^{(0,0)(r)}\\ &=&\frac {(2\pi
)^{(\frac{1}{4})}}{4\pi^{2}}\lambda \tanh (\frac {\pi \lambda
}{2})P_{-\frac{1}{2}(1+i \lambda )}(\cosh (2r))\nonumber,
\end{eqnarray}
and $P_{\nu}$ denote the Legendre function of the first kind with
parameter $\nu$.
\end{Pro}
{\bf Proof.} Let $f$ be an element  of
${\mathcal{C}}_{com}^{\infty}(D)$, the equality (1.8)' in
combination with (1.8)'' gives
\begin{eqnarray}
&&\\ I\!\!P_{\lambda}f(z)&=&\frac{1}{4\pi}\lambda \tanh (\frac
{\pi \lambda
}{2})\int_{S^1}{\cal{P}}_{\lambda}(z,w)[\int_{D}f(z'){\cal{P}}_{-\lambda}(z',w)\,dz']\,d\sigma(w).\nonumber
\end{eqnarray}
According to the Fubini theorem, the above equality becomes
\begin{eqnarray}
&&\\ I\!\!P_{\lambda}f(z)&=&\frac{1}{4\pi}\lambda \tanh (\frac
{\pi \lambda
}{2})\int_{D}f(z')[\int_{S^1}{\cal{P}}_{\lambda}(z,w){\cal{P}}_{-\lambda}(z',w)\,dz']\,d\sigma(
w)]dz'\nonumber\\
&=&\int_{D}\varphi_{\lambda}(d(z,z'))f(z')\,dz'\nonumber,
\end{eqnarray}
where $d(z,z')$ denotes the distance from $z$ to $z'$, and
$\varphi_{\lambda}$ is a multiple of the usual spherical function,
because $\varphi_{\lambda} (0)= \frac{1}{2}\lambda \tanh(\frac{\pi
\lambda}{2})$. The basic formula  for $\varphi_{\lambda}$ is
\begin{eqnarray}
\varphi_{\lambda}(d(z,z'))=\frac{1}{4\pi}\lambda \tanh (\frac {\pi
\lambda
}{2})\int_{S^1}{\cal{P}}_{\lambda}(z,w){\cal{P}}_{-\lambda}(z',w)\,d\sigma(w),
\end{eqnarray}
by taking $z'=(0,0)$ and $z=\tanh r e^{i\theta}$ to get
\begin{eqnarray}
\varphi_{\lambda}(tanh(r))=\frac{1}{4\pi}\lambda \tanh(\frac{\pi
\lambda}{2})\int_{S^1}{\mathcal{P}}_{\lambda}(\tanh r
e^{i\phi},w)\,d\sigma(w),
\end{eqnarray}
if we substtitute $w=e^{i\theta}$, we obtain ( cf. [20] p: 38)
\begin{eqnarray}
&&\\ \varphi_{\lambda}(\tanh(r))&=&\frac{1}{4\pi^2}\lambda
\tanh(\frac{\pi \lambda}{2})\int_{0}^{\pi}(\cosh
(2r)-\sinh(2r)\cos \theta)^{-\frac{1}{2}(i\lambda +1)}\,d
\theta.\nonumber
\end{eqnarray}
In his article ( cf. [26] p: 80  formula 4.5) R. Strichartz shows
that
\begin{eqnarray}
&&\\ \int_{0}^{\pi}(\cosh (2r)-\sinh(2r)\cos
\theta)^{-\frac{1}{2}+\frac{1}{2}i\lambda}\,d
\theta&=&(2\pi)^{\frac{1}{4}}P_{-\frac{1}{2}+\frac{1}{2}i\lambda}(\cosh
2r)\nonumber\\ &=&(2\pi)^{\frac{1}{4}}P_{-\frac{1}{2}(i\lambda
+1)}^{0}(2\cosh r)\nonumber
\end{eqnarray}
where $P_{\mu}^{\nu}$ denotes the Legendre functions.
\begin{eqnarray}
&&\\ P_{\mu}^{\nu}(\cosh r)&=&\frac{2^{\nu}}{\Gamma(1-\nu)}(\sinh
r)^{-\nu}\,\, _{2}F_{1}(1-\nu +\mu,-\nu
-\mu;1-\nu;\frac{1}{2}(1-\cosh(2r))).\nonumber
\end{eqnarray}
Whence
\begin{eqnarray*}
P_{-\frac{1}{2}(i\lambda +1)}^{0}(\cosh
r)&=&_{2}F_{1}(\frac{1+i\lambda}{2},\frac{1-i\lambda}{2};1;\frac{1}{2}(1-\cosh(2r)))\\
&=&_{2}F_{1}(\frac{1+i\lambda}{2},\frac{1-i\lambda}{2};1;-\sinh^{2}r)\\
&=&_{2}F_{1}(\frac{1+i\lambda}{2},\frac{1-i\lambda}{2};1;\frac{\tanh^2
r}{\tanh^2-1})\\ &=&(1-\tanh^2 r)^{\frac{1+i\lambda}{2}}
\,\,_{2}F_{1}(\frac{1+i\lambda}{2},\frac{1-i\lambda}{2};1;\tanh^2
r),
\end{eqnarray*}
since
$_{2}F_{1}(a,b;c;z)=(1-z)^{-a}\,\,_{2}F_{1}(a,c-b;c;\frac{z}{z-1}).$\\
\\
and $\phi_{\lambda}^{(a,b)}$ denote the Jacobi function
\begin{eqnarray}
&&\\
\phi_{\lambda}^{(a,b)}&=&F(\frac{a+b+1+i\lambda}{2},\frac{a+b+1-i\lambda}{2};a+1;-\sinh^2(r)).\nonumber
\end{eqnarray}
Now we give an expression exhibits $I\!\!P_{\lambda}f(z)$ as a
meromorphic function of $\lambda$, making appear zeros and poles.
\begin{Pro}
If $f$ is an element of
${\mathcal{C}}_{com}^{\infty}(\overline{B_{R}(z_0)})$,($f$ be
assumed of the form $f_{n}(\tanh(r))e^{i n \theta}$, ($n\in
l\!\!Z$)) then
\begin{eqnarray}
&&\\ I\!\!P_{\lambda}f(z)&=&\gamma(\lambda,n)(\tanh
r)^{|n|}\phi_{\lambda}^{(|n|,-|n|)}(r)e^{in\theta}\,\,.\,\,\int_{0}^{R}f_{n}(\tanh
s)\phi_{\lambda}^{(|n|,-|n|)}(s)\nonumber\\ &&\times (\tanh
s)^{|n|}\sinh(2s)\,ds \nonumber
\end{eqnarray}
with
\begin{eqnarray*}
\gamma(\lambda,n)=\frac{1}{(|n|!)^2}\frac{1}{8\pi^2}\lambda
\sinh(\frac{\pi
\lambda}{2})\Gamma(|n|+\frac{1+i\lambda}{2})\Gamma(|n|+\frac{1-i\lambda}{2}),
\end{eqnarray*}
and $f_{n}(\tanh r)$ design the Fourier's coefficient in the the
finite series of Fourier of $f$
\end{Pro}
{\bf Proof of proposition 7.2} Without loss of generality we can
assume that $z_0 = e =(0,0)$. Let $f$ be an element of
${\mathcal{C}}_{com}^{\infty}(D)$ with supp$f$ in
$(\overline{B_{R}(e)})$.\\ By combining the formulas of
$I\!\!P_{\lambda}f$ (see Remark), we see that
\begin{eqnarray}
I\!\!P_{\lambda}f(z)=\int_{S^1}{\cal{P}}_{\lambda}(z,w)\Psi_{\lambda}(w)\,d\sigma(w),
\end{eqnarray}
where
\begin{eqnarray}
\Psi_{\lambda}(w)=\frac{1}{4\pi}\lambda \tanh(\frac{\pi
\lambda}{2})\int_{D}f(z'){\cal{P}}_{-\lambda}(z',w)\,dz'.
\end{eqnarray}
Recall that the hyperbolic area measure on $D$ in geodesic polar
coordinate is given by
\begin{eqnarray*}
\frac{1}{2}\sinh(2r)dr\,d\sigma(\theta)=
\frac{1}{4\pi}\sinh(2r)dr\,d\theta.
\end{eqnarray*}
In such coordinates, the formula (7.15) becomes
\begin{eqnarray*}
\Psi_{\lambda}(e^{i\varphi})&=&\frac{1}{4\pi}\lambda
\tanh(\frac{\pi \lambda}{2})\int_{0}^{R}\int_{S^1}f_n(\tanh
r'){\cal{P}}_{-\lambda}(\tanh r'e^{i\theta},e^{i\varphi})\\
&&\times
e^{in\theta'}\,\frac{1}{2}\sinh(2r')dr'\,d\sigma(\theta')\\
&=&\frac{1}{8\pi}\int_{0}^{R}f_n(\tanh
r')[\int_{S^1}{\cal{P}}_{-\lambda}(\tanh
r'e^{i\theta},e^{i\varphi})\\ &&\times
e^{in\theta'}\,d\sigma(\theta')]\sinh(2r')dr'\\
&=&\frac{1}{8\pi}\int_{0}^{R}f_n(\tanh
r')\phi_{-\lambda,n}(r')e^{in\varphi}\sinh(2r')dr'\\
&=&(\frac{1}{8\pi}\int_{0}^{R}f_n(\tanh
r')\phi_{-\lambda,n}(r')\sinh(2r')dr')e^{in\varphi}\\
&=&K_{\lambda,n}(R)e^{in\varphi}.
\end{eqnarray*}
because $f$ is of the form $f_{n}(\tanh(r))e^{i n \theta}$, $(n\in
l\!\!Z)$, with $f_{n}(\tanh(r))$ a
${\mathcal{C}}^{\infty}$-function of support in $[-R,R]$. If we
substitute the above formula in formula (7.14) we obtain
\begin{eqnarray}
I\!\!P_{\lambda}f(z)&=&K_{\lambda,n}(R)\int_{S^1}{\cal{P}}_{\lambda}(\tanh
re^{i\theta},e^{i\varphi})e^{in\varphi}\,d\sigma(\varphi)\nonumber\\
&=&K_{\lambda,n}(R)\phi_{\lambda,n}(r)e^{in\theta}\nonumber
\end{eqnarray}
According to the previous proposition, we deduce easily
\begin{eqnarray}
&&\\ I\!\!P_{\lambda}f(z)&=&\gamma(\lambda,n)(\tanh
r)^{|n|}\phi_{\lambda}^{(|n|,-|n|)}(r)e^{in
\theta}.\int_{0}^{R}f_{n}\tanh(s)\phi_{\lambda}^{(|n|,-|n|)}(s)\nonumber\\
&&\times (\tanh s)^{|n|}\sinh(2s)\,ds\nonumber
\end{eqnarray}
with
\begin{eqnarray}
&&\\ \gamma(\lambda,n)&=&\frac{1}{(|n|!)^2}\frac{1}{8\pi}\lambda
\tanh(\frac{\pi
\lambda}{2})\frac{\Gamma(|n|+\frac{1+i\lambda}{2})}{\Gamma(\frac{1+i\lambda}{2})}\frac{\Gamma(|n|+\frac{1-i\lambda}{2})}{\Gamma(\frac{1-i\lambda}{2})}\nonumber.
\end{eqnarray}
It is clear that the expressions (7.17) and (7.18) exhibits
$I\!\!P_{\lambda}f(z)$ as a meromorphic function of $\lambda$,
with poles and zeros at exactly the points where
$\gamma(\lambda,n)$ has poles and zeros, since
$\phi_{\lambda}^{(|n|,-|n|)}$ is an entire function of $\lambda$.
Recall that $\sin(\pi z)=\frac{\pi}{\Gamma(z)\Gamma(1-z)}$, then
\begin{eqnarray*}
\cosh(\pi\frac{\lambda}{2})&=&\sin(\pi \frac{1+i\lambda}{2})\\
&=&\frac{\pi}{\Gamma
(\frac{1+i\lambda}{2})\Gamma(\frac{1-i\lambda}{2})}
\end{eqnarray*}
From this, the formula (7.18) becomes
\begin{eqnarray}
&&\\ \gamma(\lambda,n)&=&\frac{1}{(|n|!)^2}\frac{1}{8\pi^2}\lambda
\sinh(\frac{\pi
\lambda}{2})\Gamma(|n|+\frac{1+i\lambda}{2})\Gamma(|n|+\frac{1-i\lambda}{2}),\nonumber
\end{eqnarray}
{\bf Remark.} We note that $I\!\!P_{\lambda}f(z)$ has a simple
poles at $\lambda_{l}=\frac{+}{}i(2k+1)$ for $k\geq |n|$, and a
simple zero at a points $\lambda_{h}=\frac{+}{}i 2 h $ for $h \in
l\!\!Z^{*}$ and a double zero at $\lambda=0$. From the formula
(7.19), we deduce that
$\frac{I\!\!P_{\lambda}f(z)}{\Gamma(|n|+\frac{1+i\lambda}{2})\Gamma(|n|+\frac{1-i\lambda}{2})}$
has an even entire expansion, since the collection of function
$\{\phi_{\lambda}^{(-|n|,|n|)}\}$ has no zeros and no poles.
\begin{Pro}
Let $f$ be an element of
${\mathcal{C}}_{com}^{\infty}(\overline{B_{R}(z_0)})$, then there
exist an invariant subspace $E'_n$ of $E_{\lambda}$ in which we
give the condition $k\geq |n|$, such that
\begin{eqnarray}
z\to Res_{\lambda=\lambda_k}I\!\!P_{\lambda}f(z)\in \oplus_{k\geq
|n|}X_k=E'_n
\end{eqnarray}
\end{Pro}
{\bf Proof.} For any function $f\in
{\mathcal{C}}_{com}^{\infty}(\overline{B_{R}(z_0)})$, the
``spectral projection" function $I\!\!P_{\lambda}f(z)$ has the
property that $$(\Delta +\lambda^2 +1)I\!\!P_{\lambda}f(z)=
0\,\,\,\,for\,\,\,\,\lambda \neq \lambda_k, $$ then
\begin{eqnarray*}
\Delta Res_{\lambda=\lambda_k}I\!\!P_{\lambda}f(z)\\ &=&\Delta
\lim_{\lambda \to \lambda_k}(\lambda
-\lambda_k)I\!\!P_{\lambda}f(z)\\ &=&\lim_{\lambda \to
\lambda_k}(\lambda -\lambda_k)\Delta  I\!\!P_{\lambda}f(z)\\
&=&-\lim_{\lambda \to \lambda_k}(\lambda
-\lambda_k)(\lambda^2+1)I\!\!P_{\lambda}f(z)\\
&=&-(\lambda_{k}^2+1)Res_{\lambda=\lambda_k}I\!\!P_{\lambda}f(z).
\end{eqnarray*}
Then, from the above remark, we must have
\begin{eqnarray}
Res_{\lambda=\lambda_k}I\!\!P_{\lambda}f(z)\in E'_n
\,\,\,\,for\,\,\,\, k\geq |n|
\end{eqnarray}
\begin{Pro} Let $f$ be of
${\mathcal{C}}_{com}^{\infty}(\overline{B_{R}(z_0)})$, then the
regular part of $I\!\!P_{\lambda}f(z)$ at $\lambda_k$ satisfy
\begin{eqnarray}
(I\!\!P_{\lambda}f(z)-(\lambda-\lambda_k)^{-1}Res_{\lambda=\lambda_k}I\!\!P_{\lambda}f(z))\vert_{\lambda=\lambda_k}\in
\widetilde{E_{\lambda}}.
\end{eqnarray}
\end{Pro}
{\bf Proof.} According to the Proposition (3.3), we deduce
\begin{eqnarray}
&&\\ (\Delta +\lambda_{k}^2
+1)(I\!\!P_{\lambda}f(z)-(\lambda-\lambda_k)^{-1}Res_{\lambda=\lambda_k}I\!\!P_{\lambda}f(z))\nonumber\\
&=&
(\lambda_k^2-\lambda^2)I\!\!P_{\lambda}f(z)\,\,\,\,for\,\,\,\,\lambda
\neq \lambda_k,\nonumber
\end{eqnarray}
 and since
$$\lim_{\lambda \to
\lambda_k}(\lambda_k^2-\lambda^2)I\!\!P_{\lambda}f(z)=-2\lambda_k
Res_{\lambda=\lambda_k}I\!\!P_{\lambda}f(z)\in E'_n.$$ Then
$$(\Delta +\lambda_{k}^2
+1)(I\!\!P_{\lambda}f(z)-(\lambda-\lambda_k)^{-1}Res_{\lambda=\lambda_k}I\!\!P_{\lambda}f(z))\vert_{\lambda=\lambda_k}\in
E'_n.$$ but
$$(I\!\!P_{\lambda}f(z)-(\lambda-\lambda_k)^{-1}Res_{\lambda=\lambda_k}I\!\!P_{\lambda}f(z))\vert_{\lambda=\lambda_k}\notin
E_{\lambda_k}.$$ Whence $$(\Delta +\lambda_{k}^2
+1)^2(I\!\!P_{\lambda}f(z)-(\lambda-\lambda_k)^{-1}Res_{\lambda=\lambda_k}I\!\!P_{\lambda}f(z))\vert_{\lambda=\lambda_k}=0,$$
More precisely
$$(I\!\!P_{\lambda}f(z)-(\lambda-\lambda_k)^{-1}Res_{\lambda=\lambda_k}I\!\!P_{\lambda}f(z))\vert_{\lambda=\lambda_k}\in
\widetilde{E_{\lambda}}.$$
\begin{center}
{\bf Proof of the theorem 7.1}
\end{center}Let $f$ be an element of
${\mathcal{C}}_{com}^{\infty}(\overline{B_{R}(z_0)})$,  assumed
$SO(2)$- finite, then $f$ is written as a finite sum of the form
$f_m(\tanh (r)e^{im\theta}$ (with $m\in l\!\!Z$). It suffices to
verify the conditions $1),...,5)$ of theorem, for a functions of
the form  $f_m(\tanh (r)e^{im\theta}$.\\ The condition $1)$ and
$2)$ are easy, while $3)$ and $5)$ has been already established (
see propositions 7.3 and 7.4 and formula (7.18)).\\ Now showing
the estimate in $4)$. From the proposition 7.1, we have
\begin{eqnarray}
I\!\!P_{\lambda}f(z)=\int_{D}\varphi_{\lambda}(d(z,z'))f(z')\,dz',
\end{eqnarray}
where
\begin{eqnarray}
\varphi_{\lambda}(\tanh r)=\frac {(2\pi
)^{(\frac{1}{4})}}{4\pi^{2}}\lambda \tanh (\frac {\pi \lambda
}{2})P_{-\frac{1}{2}(1+i \lambda )}(\cosh (2r)),
\end{eqnarray}
For simplicity of notation we take $z=e=(0,0)$, and write
\begin{eqnarray}
F(r)=\int_{0}^{2\pi}f(tanh r e^{i\theta})\,d\theta.
\end{eqnarray}
Note that $F\in
{\mathcal{C}}_{com}^{\infty}([-R-d(z,z_0),R+d(z,z_0)])$.\\ The
formula (7.23) becomes
\begin{eqnarray}
&&\\
I\!\!P_{\lambda}f(z)&=&\frac{1}{4\pi}\int_{0}^{R+d(z,z_0)}\varphi_{\lambda}(\tanh
r)F(r)\sinh(2r)\,dr\nonumber\\ &=&\frac {(2\pi
)^{(\frac{1}{4})}}{2^{4}\pi^{3}}\lambda \tanh (\frac {\pi \lambda
}{2})\int_{0}^{R+d(z,z_0)}P_{-\frac{1}{2}(1+i \lambda )}(\cosh
(2r))F(r)\sinh(2r)\,dr.\nonumber
\end{eqnarray}
Now, we use the well-known identity (cf. [26] p: 87)  for Legendre
function $$P_{-\frac{1}{2}(1+i \lambda )}(\cosh
(r))=\frac{\sqrt{2}}{\pi}\int_{0}^{r}\frac{\cos(\lambda t)}{(\cosh
r -\cosh t)^{\frac{1}{2}}}\,dt.$$ We interchange the order of
integration in the formula (7. 26) to obtain that
\begin{eqnarray*}
\int_{0}^{R+d(z,z_0)}P_{-\frac{1}{2}(1+i \lambda )}(\cosh
(2r))F(r)\sinh(2r)\,dr=\frac{\sqrt{2}}{\pi}\int_{0}^{r}\cos(\lambda
t)G(t)\,dt,
\end{eqnarray*}
where
\begin{eqnarray*}
G(t)=\int_{t}^{R+d(z,z_0)}\frac{F(r)\sinh(2r)}{(\cosh 2r -\cosh
2t)^{\frac{1}{2}}}\,dr.
\end{eqnarray*}
To see that $G$ is $\cal{C}^{\infty}$, we note that
$(\frac{1}{\sinh r}\frac{\partial}{\partial r})^{k}F(r)$ is
$\cal{C}^{\infty}$ for any $k$, so an integration by parts yields
$$G(t)=\frac{(-1)^k}{(\frac{1}{2})_k}\int_{0}^{R+d(z,z_0)}(\cosh
(2r)-\sinh(2r))^{k-\frac{1}{2}}\sinh(2r)(\frac{d}{d\cosh (2r)})^k
F(r)\,dr,$$ is ${\mathcal{C}}^{k-1}$ by inspection.\\
\\
It remains to verify that $$\sum_{k\in
1\!\!Z}Res_{\lambda=\lambda_k}I\!\!P_{\lambda}f=0.$$ from the
inversion formula, we have (see the above proposition )
$$f(z)=\int_{I\!\!R}I\!\!P_{\lambda}f(z),$$with
\begin{eqnarray*}
I\!\!P_{\lambda}f(z)&=&\gamma(\lambda,n)(\tanh
r)^{|n|}\phi_{\lambda}^{(|n|,-|n|)}(r)e^{in\theta}.\int_{0}^{R}f_{n}(\tanh
s)\phi_{\lambda}^{(|n|,-|n|)}(s)\\ &&\times(\tanh
s)^{|n|}\sinh(2s)\,ds
\end{eqnarray*}
with
\begin{eqnarray*}
\gamma(\lambda,n)=\frac{1}{(|n|!)^2}\frac{1}{8\pi^2}\lambda
\sinh(\frac{\pi
\lambda}{2})\Gamma(|n|+\frac{1+i\lambda}{2})\Gamma(|n|+\frac{1-i\lambda}{2}).
\end{eqnarray*}
As $I\!\!P_{\lambda}f(z)$ is a meromorphic function, with poles at
$\lambda_{k}=\frac{+}{}i(2k+1)$, thus, we can use the residue
theorem to change the path of integration to obtain
$$f(z)=\int_{I\!\!R+i\alpha_k}I\!\!P_{\lambda}f(z)\,d\lambda +
2i\pi \sum_{j\leq k}Res_{\lambda=\lambda_j}I\!\!P_{\lambda}f(z),$$
where $\alpha_{k}=i(2k+1+\frac{1}{2})$, for any integer $k$. Since
$f(z)=0$ for $z\notin B_R(z_0)$, so $r\geq R$, then, it follows
from the estimate in $4)$ that $$\lim_{\alpha_{k}\to
\infty}\int_{I\!\!R+i\alpha_k}I\!\!P_{\lambda}f(z)\,d\lambda=0. $$
Whence $$\sum_{j\leq
k}Res_{\lambda=\lambda_j}I\!\!P_{\lambda}f(z)=0\,\,\,\,for\,\,\,
z\notin B_R(z_0)$$ since the above equality is a finite sum of an
fonction propre de $\Delta$, then it is a real-analytic function,
so it vanishes for all $z\in l\!\!\!C$.
\begin{Teo}
Let $F(\lambda,z)$ a function given satisfying the following
assertions\\
\\
1)\hspace{0.5cm} $F(\lambda,z)$ is $C^{\infty}$ function on
$(l\!\!\!C-il\!\!\!Z)\times D$\\
\\
2)$\hspace{0.5cm}$for each fixed $\lambda \in l\!\!\!C-il\!\!\!Z$,
we have  $\Delta_{D} F(\lambda,z)=-(\lambda^{2}+1)F(\lambda,z)$\\
\\
3)$\hspace{0.5cm}$for each fixed $z$, $F(\lambda,z)$ is an even
function meromorphic function of $\lambda$ with at worst simple
poles at $\lambda_{k}=\frac{+}{}i(2k+1)$, and
\begin{eqnarray}
\sum_{k\in l\!\!\!Z}Res_{\lambda=\lambda_{k}}F(\lambda,z)=0
\end{eqnarray}
4)$\hspace{0.5cm}$for every $N$ there exists $c_N$ such that
\begin{eqnarray}
|F(\lambda,z)|\leq c_{N}(1+|\lambda| )^{-N}e^{(R+d(z,z_{0}))|Im
\lambda|}
\end{eqnarray}
5)\hspace{0.5cm}$F(\lambda,z)$ has a simple zeros at points
$\lambda_l=\frac{+}{} i 2l$ ($l\in l\!\!Z^{*})$ and a double zero
at $\lambda=0$ and  satisfies\\
\\
$\bullet$ $z\to Res_{\lambda=\lambda_{k}}F(\lambda,z)\in E'_{n}$\\
\\
$\bullet$ $(F(\lambda,z)-(\lambda
-\lambda_{k})^{-1}Res_{\lambda=\lambda_{k}})\vert_{\lambda=\lambda_{k}}
\in \widetilde{E}_{\lambda}$\\
\\
$\bullet$ $F(\lambda,z)\vert_{\lambda=0}=0$ and
$\frac{F(\lambda,z)}{\Gamma(|k|+\frac{1+i\lambda}{2})\Gamma(|k|+\frac{1-i\lambda}{2})}$
has even entire expansion. then there exists $f$,
${\mathcal{C}}^{\infty}$ with support in $\overline{B_{R}(z_{0})}$
given by $f(z)=\int_{I\!\!R}F(\lambda,z)\,d\lambda$ such that
$F(\lambda,z)=I\!\!P_{\lambda}f(z)$
\end{Teo}
\begin{center}
{\bf Proof of theorem 7.2.}
\end{center}
Let $F(\lambda,z)$ given satisfying $1),...,5)$, and define
\begin{eqnarray}
f(z)=\int_{I\!\!R}F(\lambda,z)\,d\lambda,
\end{eqnarray}
 there is no difficulty with convergence in view of the following inequality $$\int_{I\!\!R}F(\lambda,z)\,d\lambda\leq c_N\int_{I\!\!R}(1+|\lambda|)^{-N}\,d\lambda<\infty.$$
Firt, we want to show that $f$ vanishes outside $B_R(z_0)$ and $f$
is $\cal{C}^{\infty}$.\\ For this, we assume all the functions are
of the form $f_n(\tanh r)e^{i n \theta}$ (since the condition
$1),...,5)$ are preserved), then, we can assume
\begin{eqnarray}
F(\lambda, \tanh (r) e^{i
\theta})=\Psi(\lambda)e^{in\theta}(\tanh
r)^{|n|}\phi_{\lambda}^{(|n|,-|n|)}(r),
\end{eqnarray}
for some meromorphic function, from the estimate in $4)$ we have
$$\Psi(\lambda)e^{in\theta}(\tanh
r)^{|n|}\phi_{\lambda}^{(|n|,-|n|)}(r)\leq
c_N(1+|\lambda|)^{-N}e^{(R+r)|Im \lambda|}.$$ Using the
koornwinder lemma to obtain
\begin{eqnarray}
\Psi(\lambda)\leq c_N(1+|\lambda|)^{-N}e^{r|Im \lambda|}.
\end{eqnarray}
We note that $\Psi$ is given by the condition $5).$\\ According to
the condition $3)$, we use the residue theorem in the formula
(7.26) to obtain
\begin{eqnarray*}
f(z)&=&\int_{I\!\!R}F(\lambda,z)\,d\lambda\\
&=&\int_{I\!\!R+i\alpha_k}F(\lambda,z)\,d\lambda + 2i\pi
\sum_{j\leq k}Res_{\lambda=\lambda_j}F(\lambda,z)\\ &=&
\int_{I\!\!R+i\alpha_k}F(\lambda,z)\,d\lambda
\end{eqnarray*}
where $\alpha_{k}=i(2k+1+\frac{1}{2})$, for any integer $k$. From
the formula (7.28), we conclude that
\begin{eqnarray}
&&\\ \int_{I\!\!R+i\alpha_k}F(\lambda,z)\,d\lambda&=&e^{i n
\theta}(\tanh
r)^{|n|}\int_{I\!\!R+i\alpha_k}\Psi(\lambda)\phi_{\lambda}^{(|n|,-|n|)}(r).\nonumber
\end{eqnarray}
Since $\Psi(\lambda)$ satisfy the inequality (42), then, we let
$k\to \infty$, we obtain zero in the limit in the formula (43) if
$r\geq R$. Then $f(\tanh r e^{i\theta})=0$ if $r\geq R$.\\
\\
Show now that $f$ is $\cal{C}^{\infty}$.\\
\\
Since $$\Delta^N F(\lambda,z)=(-1)^{N}(1+\lambda^2)^N
F(\lambda,z)\,\,\,for\,\,\,any\,\,\,integer\,\,\,N. $$ Then
$$\Delta^N f(z)=(-1)^{N}\int_{I\!\!R}(1+\lambda^2)^N
F(\lambda,z)\,d\lambda.$$ From the estimate in $4)$ it follows
that $\Delta^N f$ is $L^{2}(D,dm(z))$, the usual Sobolev
inequality imply that $\Delta^N f$ is $\cal{C}^{\infty}$ on $D$.\\
\\
To understand the behavior of $F(\lambda,z)$ as $\lambda \to
i(2l)$ ($l\in l\!\!Z$), we note that
$g_{z}(\lambda)=\frac{F(\lambda,z)}{\Gamma(|n|+\frac{1+i\lambda}{2})\Gamma(|n|+\frac{1-i\lambda}{2})}$
has an even entire expansion, satisfying  the estimate in $4)$,
then there is a constant $c'_N$ such that $$g_{z}(\lambda)\leq
c'_N(1+|\lambda|)^{-N}e^{(R+d(z_0,z))|Im\lambda|}.$$  From the
above results the function $\lambda \to
\Gamma(|n|+\frac{1+i\lambda}{2})\Gamma(|n|+\frac{1-i\lambda}{2})f_{\lambda,n}(\tanh
r)$ is entire and satisfying the estimate in $4)$. Using
$$f_{\lambda,n}(\tanh r)=c'_{N,n}g_{\tanh
r}(\lambda)\Gamma(|n|+\frac{1+i\lambda}{2})\Gamma(|n|+\frac{1-i\lambda}{2})|\tanh
r|^{|n|}\phi_{\lambda}^{|n|,-|n|}(r),$$ which show that $g_{\tanh
r}(\lambda)$ is an even and entire function since the collection
of function $\{\phi_{\lambda}^{(-|n|,|n|)}\}$ has no zeros and no
poles.
\\
To complete the proof we need to show that
$F(\lambda,z)=I\!\!P_{\lambda}f(z)$ which is equivalent to showing
that $\int_{I\!\!R}F(\lambda,z)=0$ imply $F(\lambda,z)=0$, and it
suffices to show that $\int_{I\!\!R}f_{\lambda,n}=0$ imply
$f_{\lambda,n}=0$. From the formula (7.27) we have
\begin{eqnarray*}
0=\int_{I\!\!R}f_{\lambda,n}(z)\,d\lambda=|\tanh
r|^{|n|}\int_{I\!\!R}\Psi(\lambda)\phi_{\lambda}^{(|n|,-|n|)}(r)\,d\lambda.
\end{eqnarray*}
Since $\Psi(\lambda)$ has compact support, by the uniqueness of
the jacobi transform (cf. [22]), we deduce that $\Psi(\lambda)=0$
for $\lambda \neq \frac{+}{}i(2k+1)$. Then $F(\lambda,z)=0$ for
$\lambda \neq \frac{+}{}i(2k+1)$.\\ {\bf acknowledgements:} The
authors would like to thank the Referee for his (her) comments and
suggestions.

\end{document}